\documentclass[a4]{article}
\usepackage{full page}
\makeatletter{}\usepackage{authblk}

\usepackage{times}
\usepackage{helvet}
\usepackage{courier}

\usepackage{flafter}
\usepackage{graphicx}
\graphicspath{{figs/}}
\usepackage{colortbl} \definecolor{grey}{rgb}{0.95,0.95,0.95}
\definecolor{darkgrey}{rgb}{0.6,0.6,0.6}

\usepackage[T1]{fontenc}
\usepackage[utf8]{inputenc}
\fontfamily{ptm}\selectfont

\usepackage{longtable}
\usepackage{float}
\usepackage{amsmath}
\usepackage{amsfonts}
\usepackage{amssymb}
\usepackage{amscd}
\usepackage{bbm}
\usepackage{easybmat}

\usepackage{algorithm}
\usepackage{algorithmic}
\usepackage{mathrsfs}
\usepackage{wrapfig}
\usepackage{paralist}
\usepackage{url}
\usepackage{xspace}
\usepackage{rotating}
\usepackage{multirow}

\usepackage{subfig}
\usepackage[utf8]{inputenc}
\usepackage{xspace}
\usepackage{bbm}
\usepackage{url}

\makeatletter{}\newcommand{\eg}{e.\,g.,\xspace}
\newcommand{\ie}{i.~e.,\xspace}
\newcommand{\cf}{cf.\xspace}

\newcommand{\flickr}{flickr\xspace}
\newcommand{\bibs}{BibSonomy\xspace}
\newcommand{\twitter}{Twitter\xspace}

\newcommand{\nameling}{Nameling\xspace}
\newcommand{\wikipedia}{Wikipedia\xspace}
\newcommand{\wiktionary}{Wiktionary\xspace}

\newcommand{\mymedia}{myMediaLite\xspace}
\newcommand{\mahout}{\ensuremath{\text{Mahout}}\xspace}
\newcommand{\eigen}{\ensuremath{\text{eigen}}\xspace}

\newcommand{\enter}[1][]{\emph{Enter#1}\xspace}
\newcommand{\click}{\emph{Click}\xspace}
\newcommand{\favorite}{\emph{Favorite}\xspace}
\newcommand{\test}[1][]{\emph{Test#1}\xspace}

\newcommand{\tl}[1]{\begin{turn}{90}#1\end{turn}}

\newcommand{\comments}[1]{}
\newcommand{\todo}[1]{}

\newcommand{\qap}{QAP\xspace}

\newcommand{\precision}[1][]{\ensuremath{\text{Precision#1}}\xspace}
\newcommand{\recall}[1][]{\ensuremath{\text{Recall#1}}\xspace}
\newcommand{\SIM}{\ensuremath{\mathop{S\!I\!M}}\xspace}
\newcommand{\ndcg}[1][]{\ensuremath{\text{NDCG#1}}\xspace}
\newcommand{\idcg}[1][]{\ensuremath{\text{IDCG#1}}\xspace}
\newcommand{\dcg}[1][]{\ensuremath{\text{DCG#1}}\xspace}
\newcommand{\avep}{\ensuremath{\text{AveP}}\xspace}
\newcommand{\map}{\ensuremath{\text{MAP}}\xspace}

\newcommand{\cov}{\ensuremath{\mathop{cov}}}
\newcommand{\var}{\ensuremath{\mathop{var}}}

\newcommand{\CN}{\ensuremath{\mathop{C\!N}}\xspace}
\newcommand{\wCN}{\ensuremath{\widetilde{\mathop{C\!N}}}\xspace}
\newcommand{\JC}{\ensuremath{\mathop{J\!AC}}\xspace}
\newcommand{\wJC}{\ensuremath{\widetilde{\mathop{J\!AC}}}\xspace}
\newcommand{\PA}{\ensuremath{\mathop{P\!A}}\xspace}
\newcommand{\wPA}{\ensuremath{\widetilde{\mathop{P\!A}}}\xspace}
\newcommand{\AC}{\ensuremath{\mathop{A\!A}}\xspace}
\newcommand{\wAC}{\ensuremath{\widetilde{\mathop{A\!A}}}\xspace}
\newcommand{\RA}{\ensuremath{\mathop{R\!A}}\xspace}
\newcommand{\wRA}{\ensuremath{\widetilde{\mathop{R\!A}}}\xspace}
\newcommand{\NM}{\ensuremath{\mathop{N\!EW}}\xspace}
\newcommand{\CS}{\ensuremath{\mathop{C\!O\!S}}\xspace}
\newcommand{\wCS}{\ensuremath{\widetilde{\mathop{C\!O\!S}}}\xspace}
\newcommand{\PR}{\ensuremath{\mathop{P\!R}}\xspace}
\newcommand{\PPR}{\ensuremath{\mathop{P\!P\!R}}\xspace}
\newcommand{\PPRa}{\ensuremath{\mathop{P\!P\!R\!+}}\xspace}
\newcommand{\NR}{\ensuremath{\text{NameRank}}\xspace}
\newcommand{\mostpopular}{\ensuremath{\text{MostPopular}}\xspace}
\newcommand{\MP}{\ensuremath{\text{MP}}\xspace}
\newcommand{\RND}{\ensuremath{\text{RND}}\xspace}
\newcommand{\WRMF}{\ensuremath{\text{WRMF}}\xspace}
\newcommand{\UCF}{\ensuremath{\text{UCF}}\xspace}
\newcommand{\ICF}{\ensuremath{\text{ICF}}\xspace}
\newcommand{\rec}{\ensuremath{\text{Rec}}\xspace}
\newcommand{\reco}[2]{\ensuremath{\text{Rec}^{#1}(#2)}\xspace}

\newcommand{\aveRank}{\ensuremath{\text{AveRank}}\xspace}

\newcommand{\shuffle}[1]{\ensuremath{\underline{#1}}}
\newcommand{\rewire}[1]{\ensuremath{\overline{#1}}}

\newcommand{\pagerank}{PageRank\xspace}
\newcommand{\nameRank}{\ensuremath{\mathop{P\!P\!R\!+}}\xspace}
\newcommand{\multiRank}{\ensuremath{\text{MultiRank}}\xspace}
\newcommand{\pRank}{\ensuremath{\text{PRank}}\xspace}

\newcommand{\coloneqq}{\mathrel{\mathop{:}}=}       \newcommand{\eqqcolon}{\mathrel{\mathop{=}}:}       \newcommand{\walk}[3]{\ensuremath{#1\rightarrow_{#2}#3}}

           \newcommand{\R}{\mathbb{R}}

\newcommand{\size}[1]{\ensuremath{|#1|}}

\newcommand{\set}[1]{\ensuremath{\{#1\}}}

\newcommand{\Nen}{\ensuremath{G_{\text{\textsc{EN}}}}\xspace}
\newcommand{\Nde}{\ensuremath{G_{\text{\textsc{DE}}}}\xspace}
\newcommand{\Nfr}{\ensuremath{G_{\text{\textsc{FR}}}}\xspace}

\newcommand{\wikiDE}{\ensuremath{\text{Wiki}^{\text{DE}}}\xspace}
\newcommand{\wikiEN}{\ensuremath{\text{Wiki}^{\text{EN}}}\xspace}

\newcommand{\Cen}{\ensuremath{G^C_{\text{\textsc{EN}}}}\xspace}
\newcommand{\Cde}{\ensuremath{G^C_{\text{\textsc{DE}}}}\xspace}
\newcommand{\Cfr}{\ensuremath{G^C_{\text{\textsc{FR}}}}\xspace}

\newcommand{\CTweets}{\ensuremath{G^C_{\text{Twitter}}}\xspace}
\newcommand{\NTweets}{\ensuremath{G_{\text{Twitter}}}\xspace}

\newcommand{\Group}{Group}
\newcommand{\Friend}{Friend}

\newcommand{\cooc}{co-occurrence}

\DeclareCaptionType{copyrightbox}
\begin{document}

            \makeatletter{}\title{Recommending Given Names\\[0.25\baselineskip]
\large{Mining Relatedness of Given Names based on Data from the
Social Web}}
 \date{}

\author[1]{Folke Mitzlaff}
\author[1]{Gerd Stumme}
\affil[1]{ University of Kassel\\
              Knowledge and Data Engineering Group\\
              Kassel, Germany}

   \maketitle

            \makeatletter{}\begin{abstract}
\comments{
\begin{verbatim}
* present task: name recommendations
* present nameling
* aggregate and analyse evidenz 
  networks of user relatedness
* set up collaborative filtering based 
  recommenders
* set up evaluation scenario
* compare to static recommender 
  (cosine/pagerank)
\end{verbatim}}
  All over the world, future parents are facing the task of finding a
  suitable given name for their child. This choice is influenced by
  different factors, such as the social context, language, cultural
  background and especially personal taste. Although this task is
  omnipresent, little research has been conducted on the analysis and
  application of interrelations among given names from a data mining
  perspective.

  The present work tackles the problem of recommending given names, by
  firstly mining for inter-name relatedness in data from the Social
  Web. Based on these results, the name search engine
  ``\nameling''\footnote{\url{http://nameling.net}} was built, which
  attracted more than 35,000 users within less than six months,
  underpinning the relevance of the underlying recommendation
  task. The accruing usage data is then used for evaluating different
  state-of-the-art recommendation systems, as well our new \NR
  algorithm which we adopted from our previous work on folksonomies
  and which yields the best results, considering the trade-off between
  prediction accuracy and runtime performance as well as its ability
  to generate \emph{personalized} recommendations. We also show, how
  the gathered inter-name relationships can be used for meaningful
  result diversification of PageRank-based recommendation systems.

  As all of the considered usage data is made publicly
  available\footnote{\url{http://www.kde.cs.uni-kassel.de/nameling/dumps}},
  the present work establishes baseline results, encouraging other
  researchers to implement advanced recommendation systems for given
  names.
\end{abstract}
 
      \makeatletter{}\section{Introduction}\label{sec:introduction}
The choice of a given name is typically accompanied with an extensive
search for the most suitable alternatives, at which many constraints
apply. First of all, the social and cultural background determines,
what a common name is and may additional imply certain habits, such
as, \eg the patronym. Additionally, most names bear a certain meaning
or associations which, also depend on the cultural context.

Whoever makes the decision is strongly influenced by
personal taste and current trends within the social context. Either by
preferring names which are currently popular, or by avoiding names
which most likely will be common in the neighborhood. Recently, public
discussion of psychological effects associated with certain given
names, which eventually may lead to social discrimination of
individuals~\cite{nelson2007moniker}, increased the perception of
responsibility of future parents, making the process of finding a
given name even more involved.

Future parents are often aided by huge collections of given names
which list several thousand names, ordered alphabetically or by
popularity. With the first author's need for a given name for his
child, the idea arose to collect data from the ``Social Web'' in order
to derive background information, popularity, interrelations and
similarities of given names. The search engine
``\nameling''~\cite{mitzlaff2012namelings} utilizes \wikipedia's text
corpus for interlinking names and the micro-blogging service \twitter
for capturing current trends and popularity of given
names. Nevertheless, the underlying
rankings~\cite{mitzlaff2012ranking} and thus the search results are
statically bound to the underlying co-occurrence graph obtained from
\wikipedia and thus not personalized.

This work presents the task of recommending given names, applying
approaches for distributional semantics and state-of-the-art
recommender systems, such as user based collaborative filtering
(\UCF), item based collaborative filtering (\ICF) and matrix
factorization approaches. Additionally, the recommendation algorithm
\NR is presented, which is adopted from previous work on
folksonomies~\cite{hotho2006information}, showing good performance in
terms of prediction accuracy as well as runtime complexity.

The rest of the work is structured as follows:
In Sec.~\ref{names:relatedness}, inter-name similarities are derived
and evaluated. Sec.~\ref{names:recommendation} presents the
usage data of the \nameling search engine which is then used for
evaluating name recommendation approaches. Then, in
Sec.~\ref{sec:diversification},  different approaches for result
diversification are discussed and evaluated. \comments{Sec.~\ref{sec:related}
presents the overall related work.}

            \makeatletter{}\section{Relatedness of Given Names}\label{names:relatedness}
\makeatletter{}\label{names:intro}
\comments{
  AIM:
  interrelations among given names -> use for search/ranking/recommendation

  many reasons for names to be related
  - phoentics
  - ``meaning''
  - popularh older of the name (ie history/royal
  family/literature/movies/celebrities...)
  - common patterns (ie max&moritz/...)

  rise of social web: whelm of data available which interconnects
  people/objects/information

  ==> use this datasource for mining object inerrelatedness (HT13)

  as manifaceted names and their interrelationship -> as manifaceted
  users/individuals/... interconnect 

  ultimately: user/social context/name/cultural context/sentiment
  mining...

  here: first, simple, (co-oc) yet powerful approach (success of
  nameling)
}
This section aims at discovering relations among given names, which
can be used to aid future parents in searching and finding suitable
names by means of ranking and recommendations techniques.

There are many reasons for names being related, ranging from phonetic
similarity to common cultural context or similar ``meaning''. With the
rise of the Social Web, a whelm of data sources becomes available,
interconnecting users and object information, either explicitly or
implicitly~\cite{mitzlaff2013semantics}. In particular, given names
are linked by social interactions of respective holder of the names as
well as occurrences within heterogeneous, mostly unstructured data
collections.

Ultimately, by leveraging data from online social networks,
microblogging systems and encyclopedic background information,
recommendation systems for given names may consider the user's social
context, cultural habits, personal taste and common popularity. In the
following, we present a simple, yet powerful approach for mining
interrelations among given names, which is based on co-occurrence
networks. 
 
\makeatletter{}\paragraph{Data Sources}\label{sec:data}
For building co-occurrences networks of given names, we used the
official \wikipedia data dump which is freely available for
download\footnote{\url{http://dumps.wikimedia.org/backup-index.html}}
and considered the English (date: \emph{2012-01-05}), French
(\emph{2012-01-17}) and German (\emph{2011-12-12}) version
separately. We additionally used the categorization of the
affiliated \wiktionary project (English, French and German
\emph{2012-06-06}), also available for download.
As an additional source for user generated data, we considered the
microblogging service \twitter. Using \twitter, each user publishes
short text messages (called ``\emph{tweets}''). We used the data set
introduced in~\cite{yang2011patterns}, which comprises 476,553,560
tweets from 17,069,982 users, collected 2009/06 until
2009/12.
Some effort was made to build up a comprehensive list of given
names. In a semi-automatic way, a list of more than 30,000 names was
collected. During the first months of the \nameling's live time,
additional names were proposed by users of the system, yielding a list
of 36,434 given names.  
 
\makeatletter{}\subsection{Networks of Given Names}\label{sec:networks}
\enlargethispage*{3\baselineskip}
One of the most basic notions of relatedness between two given names
can be observed, when they occur together in an atomic context of a
given data collection. In case of \wikipedia, we counted such
\emph{co-occurrences} based on sentences and for \twitter based on
tweets. We thus obtain for each data source
$S\in\set{\text{EN},\text{DE},\text{FR},\text{Twitter}}$ (English,
German and French \wikipedia as well as \twitter) an undirected
weighted graph $G_S=(V_S,E_S)$ where $V_S$ denotes the subset of all
observed names within $S$ and for names $u, v$ exists an edge
$(u,v)\in E_S$ with weight $w(u,v)$, if $u$ and $v$ co-occurred in
exactly $w(u,v)$ contexts.
For example, the given names ``\emph{Peter}'' and ``\emph{Paul}''
co-occurred in 30,565 sentences within the English
\wikipedia. Accordingly, there is an edge $(\text{Peter},
\text{Paul})$ in {\Nen} with corresponding edge weights.

\label{sec:statistics}
Table \ref{tab:networks:general} summarizes the high level statistics
for all considered co-occurrence networks. As one would expect, all
networks contain a giant connected
component~\cite{newman2003structure} which almost covers the whole
corresponding node sets. The network obtained from the English
\wikipedia is the most densely connected network whereas the network
obtained from \twitter is the least densely connected.
\begin{table*}
  \centering
  \caption{High level statistics for all co-occurrence networks.}
  \begin{tabular}{l|r|r|r|r|r}
             &  \multicolumn{1}{c|}{$|V|$}  
             &  \multicolumn{1}{c|}{$|E|$}         & density   &   \#wcc  & largest wcc \\\hline\hline
    \Nen     &  $27,121$ &  $24,461,988$    &  $0.067$  &   $2$    &   $27,119$  \\\hline
    \Nde     &  $25,032$ &  $13,172,603$    &  $0.042$  &   $2$    &   $25,030$  \\\hline
    \Nfr     &  $25,237$ &  $17,446,139$    &  $0.055$  &   $1$    &   $25,237$  \\\hline
    \NTweets &  $25,902$ &   $6,634,551$    &  $0.020$  &   $1$    &   $25,902$  \\
  \end{tabular}
  \label{tab:networks:general}
\end{table*}

\paragraph{Inter-Network Correlation Test}\label{sec:internetwork:qap}
For assessing the pairwise structural interdependence of the different
networks, we apply the quadradic assignment procedure (\qap)
test~\cite{butts2008social,butts2005simple}. For given graphs
$G_1=(V_1, E_1)$ and $G_2=(V_2,E_2)$ with $U\coloneqq V_1\cap
V_2\ne\emptyset$ and adjacency matrices $A_i$ corresponding to
${G_i}_{|U}$ ($G_i$ reduced to the common vertex set $U$), the graph
\emph{covariance} is given by
\[
\cov(G_1,G_2) \coloneqq
\frac{1}{n^2-1}\sum_{i=1}^{n}\sum_{j=1}^{n}(A_1[i,j]-\mu_1)(A_2[i,j]-\mu_2)
\]
where $n\coloneqq\size{U}$ and $\mu_i$ denotes $A_i$'s mean
($i=1,2$). Then $\var(G_i)\coloneqq\cov(G_i, G_i)$ leading to the
graph correlation
$
\rho(G_1,G_2)\coloneqq
\frac{\cov(G_1,G_2)}{\sqrt{\var(G_1)\var(G_2)}}.
$

The \qap test compares the observed graph correlation $\rho_0$ to the
distribution of resulting correlation scores obtained on repeated
random row/column permutations of $A_2$. The fraction of permutations
$\pi$ with correlation $\rho^\pi\ge\rho_o$ is used for assessing the
significance of an observed correlation score $\rho_o$. Intuitively,
the test determines (asymptotically) the fraction of all graphs with
the same structure as $G_{2|U}$ having at least the same level of
correlation with $G_{1|U}$.

Table \ref{tab:qap} shows the pairwise correlation scores for all
considered networks. The \wikipedia-based co-occurrence graphs shows
the strongest correlation. For assessing the significance of the
observed correlations, we repeatedly calculated the pairwise
correlations on 1,000 corresponding randomly generated null
models. For any pair of the considered networks, every randomly
generated null model showed much lower correlation scores
($<10^{-3}$), which indicates statistical
significance~\cite{butts2008social}.

We conclude that the co-occurrence networks structurally
correlate. Nevertheless, language specific deviations exist. For
discovering relations on named entities, the corresponding language
should therefore be considered. In the next section we will
investigate, how structural similarity within the co-occurrence
networks correlate with natural notions of relatedness among given
names.
\comments{Figure~\ref{fig:internetwork:qap} shows exemplary distribution
of the resulting correlation scores which are consistently lower by
magnitude then the observed correlations in the original graphs, which
suggests that even the low correlation scores for the \twitter based
networks are significant.}
\begin{table*}\centering
  \caption{
    \emph{Left}: Pairwise graph correlation observed in the co-occurrence graphs.
    \emph{Right}: Basic statistics for the different activities within the \nameling
  }
  \subfloat[][Pairwise graph correlation]{\label{tab:qap}
  \begin{minipage}{0.45\linewidth}
    \begin{tabular}{l|r|r|r|r}\scriptsize
      &  \Nen     &  \Nde      &  \Nfr     &  \NTweets \\\hline\hline
      \Nen     &  \multicolumn{1}{c|}{-}        &  $0.406$   &  $0.332$  &  $0.049$  \\\hline
      \Nde     &  \multicolumn{1}{c|}{-}        &  \multicolumn{1}{c|}{-}         &  $0.303$  &  $0.020$  \\\hline
      \Nfr     &  \multicolumn{1}{c|}{-}        &  \multicolumn{1}{c|}{-}         &  \multicolumn{1}{c|}{-}        &  $0.015$  \\
    \end{tabular}
  \end{minipage}
}\quad\quad\quad
  \subfloat[][Basic usage data statistics]{\label{tab:usage:basic}
  \begin{tabular}{l|r|r}
                   &  \#Users &  \#Names \\\hline
    \enter         & $35,684$ & $16,498$ \\\hline
    \click         & $22,339$ & $10,028$ \\\hline
    \favorite      &  $1,396$ &  $1,558$
  \end{tabular}

  }
\comments{
  \begin{minipage}{0.45\linewidth}
    \begin{tabular}{l|r|r|r|r}
      &  \Cen    &  \Cde       &  \Cfr      &  \CTweets \\\hline\hline
      \Cen     &  \multicolumn{1}{c|}{-}       &  $0.119$    &  $0.180$   &  $0.067$  \\\hline
      \Cde     &  \multicolumn{1}{c|}{-}       &  \multicolumn{1}{c|}{-}          &  $0.135$   &  $0.025$  \\\hline
      \Cfr     &  \multicolumn{1}{c|}{-}       &  \multicolumn{1}{c|}{-}          &  \multicolumn{1}{c|}{-}         &  $0.040$  \\\hline
      \CTweets &  \multicolumn{1}{c|}{-}       &  \multicolumn{1}{c|}{-}          &  \multicolumn{1}{c|}{-}         &  \multicolumn{1}{c}{-}        \\
    \end{tabular}
  \end{minipage}
}

\end{table*}

\comments{
All networks within \bibs show a consistent level
of correlation with a significant peak for the explicit \Friend- and
\Group networks. Considering the results for the networks obtained
from \flickr it is worth noting that, though low in magnitute, the
Favorite-Graph shows a significant higher correlation with the
Contact-Graph than the other pairs of networks do.
}

\comments{
\begin{figure}
  \centering
  \includegraphics[width=0.3\linewidth]{internetwork/qap/de-en-1000}
  \includegraphics[width=0.3\linewidth]{internetwork/qap/de-fr-1000}
  \includegraphics[width=0.3\linewidth]{internetwork/qap/en-fr-1000}\comments{\\
  \includegraphics[width=0.3\linewidth]{internetwork/qap/en-twitter-1000}
  \includegraphics[width=0.3\linewidth]{internetwork/qap/de-twitter-1000}
  \includegraphics[width=0.3\linewidth]{internetwork/qap/fr-twitter-1000}}
  
  \caption{Test statistics for the QAP correlation test.}
  \label{fig:internetwork:qap}
\end{figure}
} 
\makeatletter{}\subsection{Mining for Relations from the Social Web}\label{names:networks:semantics}
\label{names:task}
In this section, we focus on the question, whether structural
similarity in the co-occurrence networks from
Section~\ref{sec:networks} gives rise to a notion of relatedness which
implies relationships the user might be interested in.
For evaluating and comparing different similarity metrics, we need a ``reference''
notion of relatedness for the considered names to be used as ``ground
truth''. For given names, there is no generally accepted reference
relation. We therefore apply the approach of using an external data
source which we assume as a valid ``gold standard''. We argue that the
categories assigned to names in \wiktionary are a good basis, as they
are manually assigned and have a direct connection to concepts users
associate with given names (such as gender and cultural context). We
finally chose cosine similarity for
calculating a reference similarity score, which is broadly accepted
for various applications. For the sake of a concise presentation, we
restrict our analysis in this chapter to the network obtained from the
English \wikipedia and from \twitter.
\comments{
Firstly, we introduce different similarity functions for calculating
similarity of nodes within graphs. We will than compare these
similarity functions for given names with the corresponding gold
standard relations described above.}
\comments{future work: Of course,
  whether or not the considered similarity scores coincide with a
  concept of relatedness a user has in mind can ultimately only be
  answered by evaluating against usage statistics and user feedback in
  a running system, such as, \eg the \nameling.  }
 
\makeatletter{}\paragraph{Vertex Similarities}\label{sec:similarities}\label{sec:metrics}
Subsequently, we only consider the two similarity functions which
yielded the most promising results. For the discussion and evaluation
of broader range of similarity metrics, refer to~\cite{mitzlaff2012nameling01}. \comments{
\paragraph{Common Neighbors (\CN)}
The \CN metric simply counts common neighbors for pairs of vertices:
\[
\CN(x,y)\coloneqq\size{\Gamma(x)\cap\Gamma(y)}
\]
Although simple, the \CN is widely adopted in most online social
networks and shows good performance for predicting
links~\cite{libennowell2007linkprediction} and is straightforward
extended to weighted networks:
\[
\wCN(x,y)\coloneqq\sum_{z\in\Gamma(x)\cap\Gamma(y)}w(x,z)+w(y,z)
\]
}
The \emph{Jaccard coefficient} measures the fraction of common
neighbors. For weighted networks, we also consider its weighted
variant~\cite{murata2007prediction}.
The \emph{cosine similarity} measures the cosine of the angle between
the corresponding rows of the adjacency matrix, where we consider both
its weighted and unweighted variant.

For obtaining a reference relation on the set of given names, we
collected all corresponding category assignments from \wiktionary. We
thus obtained for each of 10,938 given names a respective binary
vector, where each component indicates whether the corresponding
category was assigned to it (in total 7,923 different categories and
80,726 non-zero entries). We laxly denote these category assignments
as ``semantic'' properties and accordingly the induced similarity as
``semantic similarity'' of names.\comments{As these assignment vectors
  are very sparse, we counted for each name the number of name pairs
  with a non-zero similarity score, to ensure that a relevant
  similarity metric is induced. Indeed, more than 90\% of the names
  had more than one hundred ``similar'' names.}

\begin{figure*}
  \centering
    \includegraphics[scale=0.65]{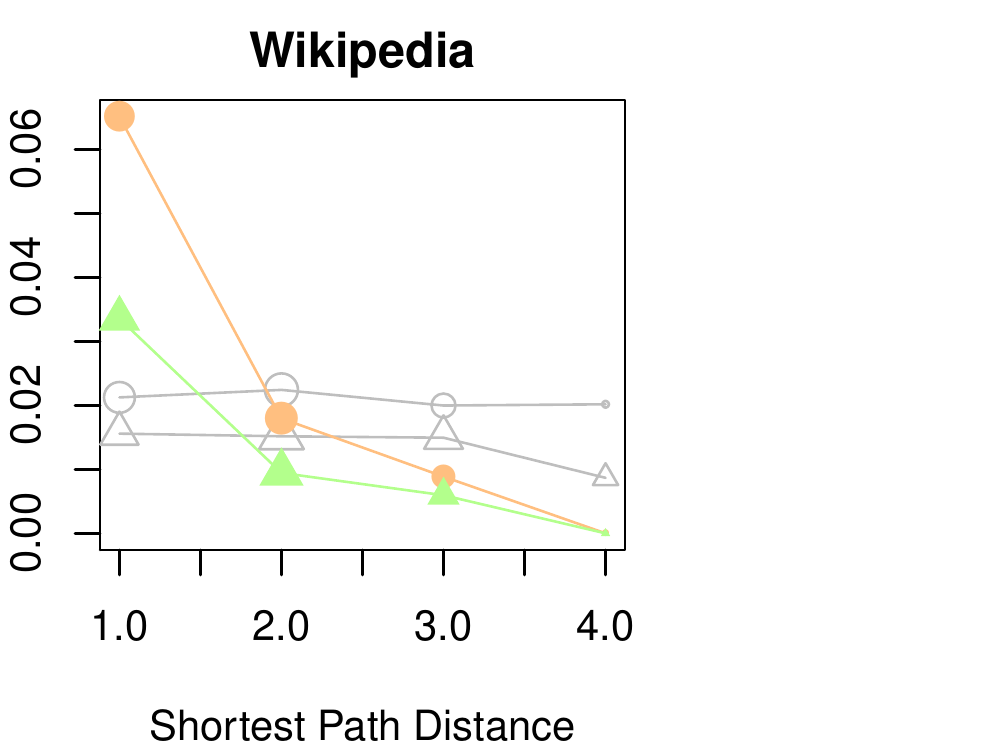}
    \includegraphics[scale=0.65]{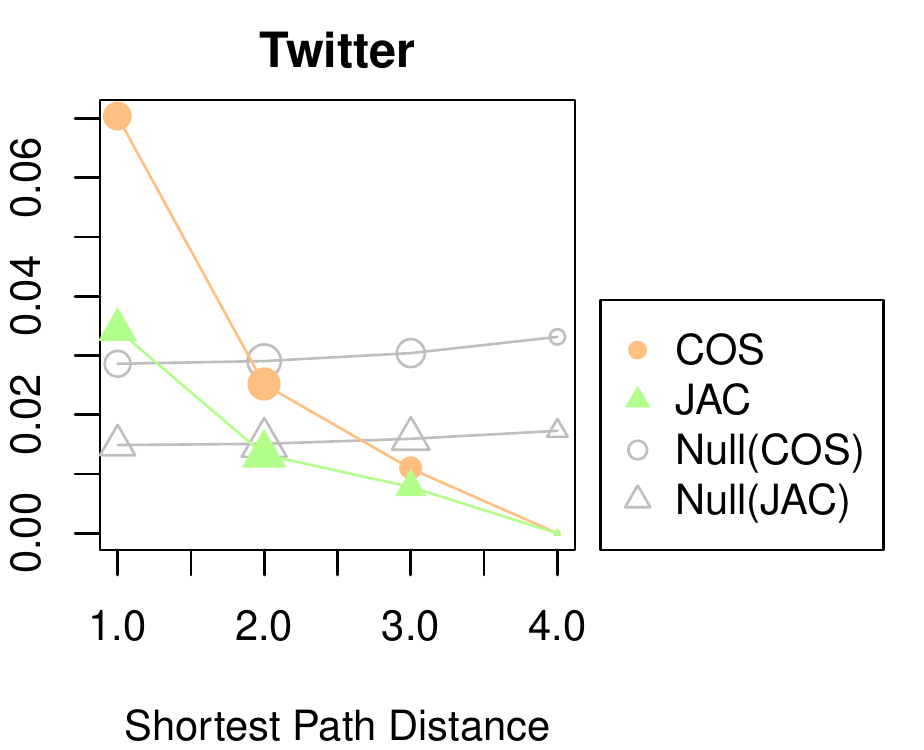}
  \caption{Semantic relatedness vs. shortest path distance in the co-occurrence networks.}
  \label{fig:similarities:simdist}
\end{figure*}

\paragraph{Neighborhood \& Similarity}\label{sec:experiments:simdist}
As a first analysis of the interdependence of a name's position within
the co-occurrence network and its category assignment in \wiktionary,
we consider for pairs of names their respective shortest path distance
relative to their reference similarity. Considering the co-occurrence
networks obtain from \wikipedia and \twitter separately, we calculate
the average corresponding similarity score for every shortest path
distance $d$ and every pair of names $u,v$ with a shortest path
distance $d$. To rule out statistical effects, the same calculations
are repeated at which the correspondence of names within the network
and the category assignment is randomly shuffled.
Fig.~\ref{fig:similarities:simdist} shows the results for the cosine
similarity together with the Jaccard coefficient. In both networks,
the similarity of node pairs tends to decrease monotonically with the
respective shortest path distance, where direct neighbors are in
average more similar than randomly chosen pairs (referring to the null
model baseline) and pairs at distance two are already less similar
than expected by chance.

\paragraph{Structural \& Semantical Similarity}\label{sec:experiments:semsim}
The interplay of shortest path distance and semantic similarity of
names indicates that the structural context of a name within the co-occurrence
network is correlated with its semantic properties.

\begin{figure*}
  \centering
  \includegraphics[scale=0.65]{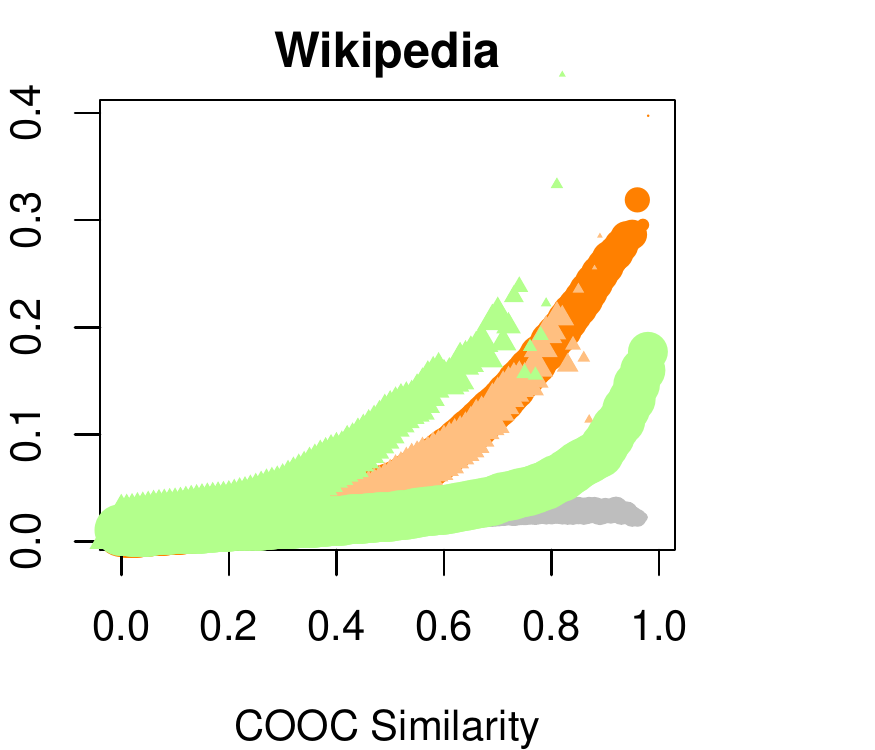}
  \includegraphics[scale=0.65]{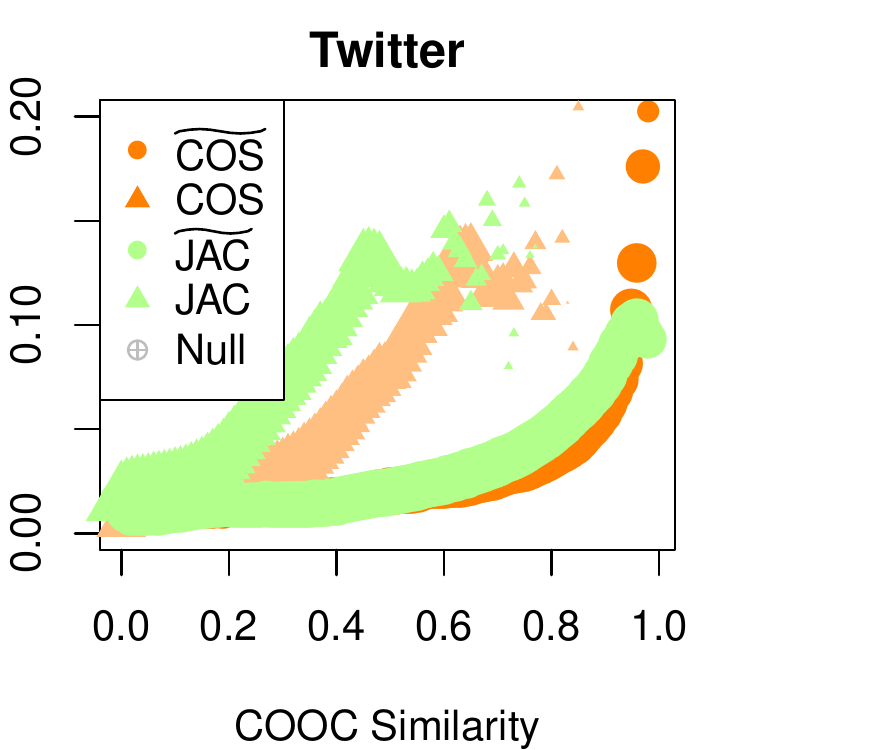}

  \caption{Similarity based on name categories in \wiktionary
    vs. vertex similarity in the \cooc-networks (weighted and
    unweighted).}
  \label{fig:similarities:semsim}
\end{figure*}

We now compare different vertex similarity metrics with respect to
their ability to capture semantic similarity of names. In detail, we
calculate for any pair $u,v$ of names in the co-occurrence network
(which have a category assignment)  the cosine similarity
$\CS(u,v)$ based on the respective category assignment vectors as well
as any of the considered vertex similarity metrics $s(u,v)$. As the
number of data points $(\CS(u,v), s(u,v))$ grows quadratically with
the number of names, we grouped the co-occurrence based similarity
scores in 1,000 equidistant bins and calculated for each bin the
average cosine similarity based on category
assignments. Figure~\ref{fig:similarities:semsim} shows the results
for \wikipedia and \twitter separately.

Notably, all considered similarity metrics capture a positive
correlation between similarity in the co-occurrence network and
similarity between category assignments to names. But significant
differences between the underlying co-occurrence networks and the
applied similarity functions can be observed. As for \wikipedia, the
cosine similarity shows similar characteristics in its weighted and
unweighted variant but for higher structural similarity scores the
weighted variant is more consistent with the semantic similarity of
names. Considering the Jaccard coefficient, the unweighted variant is
more consistent with the reference similarity then its unweighted
variant.

As for \twitter, both cosine similarity and the Jaccard coefficient
are more consistent with the reference similarity for lower and
average structural similarity scores. For higher structural
similarity, the weighted cosine similarity shows strong correlations
with the reference relation.\enlargethispage*{3\baselineskip}

Summing up, edge weights without further pre-processing may
significantly decrease the performance of a similarity metric with
respect to its ability to capture semantic similarity of names.
Furthermore, no clearly best structural similarity can be deduced from
the conducted experiments. Nevertheless, correlations between
structural and semantic similarity can be observed at all consistent
vertex similarity functions.

\section{Recommending Given Names}\label{names:recommendation}
\makeatletter{}\comments{
  previous sections: background knowledge -> relatedness -> already
  used for ranking (-> nameling => usage data)

  Now: Build personalized r. systems using this data
  -> compare different well established rs (UCF->WRMF) as well new own
  approach (item based FR)..

  1) Evaluate in terms of prediction accuracy 2) later: use/combine
  relatedness of sec ... with usage data in order to tackle the
  accuracy/diversity dilemma 
}
The results presented in the previous section indicate correlations
between semantic relatedness among given names and structural
similarity within co-occurrence networks obtained from different data
sources. 

We now turn our focus towards systems for personalized name
recommendations, leveraging the usage data which accrued at
\nameling's log files. We firstly summarize the considered usage data
and analyze the emerging networks of given names. We then briefly
summarize different state-of-the-art recommendation systems and
present our own recommendation approach.  All considered
recommendation systems are finally evaluated and compared with respect
to their prediction accuracy.
 
\makeatletter{}\subsection{Preliminaries}\label{sec:preliminaries}
Throughout this section, we utilize $u,v$ and $i,j$ interchangeably as
placeholder or index variables to denote users and items (\ie names)
respectively. We denote the set of users with $\mathcal{N}$ and the
set of items with $\mathcal{M}$, denoting the number of users with
$n\coloneqq\size{\mathcal{N}}$ and the number of names with
$m\coloneqq\size{\mathcal{M}}$.

The context of our work is given by \nameling, a search and
recommendation system for given names, where a user $u$ might express
interest in name $i$, either by entering the name directly in a search
mask, click on a name or add a name to the personal list of favorite
names. We denote $u$'s expressed affiliation with $i$ by
$r_{ui}^{\enter{}}$, $r_{ui}^\click$ and $r_{ui}^\favorite$,
respectively. All user-name affiliations are aggregated in the
corresponding binary user-name matrices $R^{\enter{}}$, $R^{\click}$
and $R^{\favorite}$. Whenever the corresponding activity class is
irrelevant or determined by the context, we drop the
superindex. Furthermore we consider the set
$\mathcal{M}_u\coloneqq\set{j\in\mathcal{M}\mid r_{uj}>0}$ of all
names $j$ which user $u$ is affiliated with as well as the set
$\mathcal{N}_i\coloneqq\set{v\in\mathcal{N}\mid r_{vi}>0}$ of all
users $v$ who are affiliated with the $i$th given name.

Please note that we consider binary user-name affiliations, \ie
$r_{ui}\in\set{0,1}$ where a zero value may either indicate that user
$u$ is not interested in $i$, has not yet expressed her/his
affiliation or is just unaware of $i$. A positive value of $r_{ui}$ on
the other hand only in the case of adding a name to the personal list
of favorites clearly indicates that $u$ \emph{likes} $i$. The reason
for entering or clicking on a name might also just be curiosity.

Recommending given names for a user $u$ based on a given user-name
matrix $R$ corresponds to predicting $u$'s affiliation with any name
$i\in\mathcal{M}\setminus\mathcal{M}_u$ which we denote with
$\hat{r}_{uj}$. Most recommendation systems determine for each such
user-item pair a score which reflects the recommender's confidence
about positive affiliation. Recommending $k$ names for $u$ is achieved
by taking the top $k$ names from the thus ordered list of names. For a
recommendation system Rec, we denote the ordered list of $k$
recommended names for user $u$ with $\reco{k}{u}$ and write
$\text{Rec}(u)$ to denote Rec's ranking of all names for $u$, based on
the recommender's scoring function.

For assessing the prediction accuracy of a recommendation system, we
process each user $u$ separately and select for evaluation a subset
$\test(u)\subset\mathcal{M}_u$. Recommendations
for $u$ are then calculated on $\tilde{R}$ which is element-wise given
by
\[
\tilde{r}_{vj}\coloneqq
\begin{cases}
  0, & \text{if } v=u \text{ and } j\in\test(u)\\
  r_{vj},& \text{otherwise.}
\end{cases}
\]
We consider several metrics for scoring a recommendation system's
prediction accuracy:
\paragraph{Precision/Recall}
Precision and recall are metrics originating from the evaluation of
information retrieval systems, given by
\begin{eqnarray*}
  \precision{^k\!(u)}&\coloneqq& \frac{\size{\reco{k}{u}\cap\test(u)}}{k}\\
  \recall{^k\!(u)}&\coloneqq& \frac{\size{\reco{k}{u}\cap\test(u)}}{\size{\test(u)}}
\end{eqnarray*}
We interchangeably also write \precision{@k} and \recall{@k} to denote
the average scores over all users $u\in\mathcal{N}$. 
\paragraph{Average Precision/MAP}
The \emph{mean average precision} is a metric for obtaining a single
value performance score for a recommendation system $\text{Rec}$
considering all recommended names for all users:
\[
\map\coloneqq \frac{1}{n}\sum_{u=1}^{n}\avep(u)
\]
where the \emph{average precision} is given by
\[
\avep(u)\coloneqq
\frac{1}{\size{\mathcal{M}_u}}\sum_{k=1}^{\size{\mathcal{M}_u}}[\precision{^k\!(u)}\cdot\delta^k(u)]
\]
and $\delta^k\!(u)=1$, iff the recommended element at rank $k$ is a
relevant name (that is $\text{Rec}(u)[k]\in\test(u)$). Refer
to~\cite{voorhees2005experiment} for more details.

\paragraph{Normalized Discounted Cumulative Gain}
The basic idea of the discounted cumulative gain metric (\dcg) is that
relevant items at higher ranking positions should be
penalized. Typically, the penalization factor scales inverse
logarithmically with the position in the list of recommended items:
\[
\dcg{^k}(u)\coloneqq \sum_{i=1}^k \frac{2^{\delta^i(u)}-1}{\log_2(i+1)}
\]
For normalizing the \dcg score across recommendations for different
users, the \dcg score is considered relative to the best possible \dcg
score for the considered user, called the \emph{ideal} score 
\idcg, resulting in 
\[
\ndcg{^k}(u)\coloneqq\frac{\dcg^k(u)}{\idcg^k(u)}
\]
For maximal $k$ we drop the index and write \ndcg. Please note hat we
restrict our discussion to the binary case.

\paragraph{Test of Significance}
For comparing the prediction accuracy of different recommendation
systems, typically the average of some evaluation metric is
compared. Due to the sensitivity of the average towards outliers such
a comparison may often lead to erroneous conclusions. For further
comparison the distribution of the respective evaluation metric should
be taken into account too, \eg in order to apply some statistical test
for assessing the significance of the observed difference.

We apply the simple \emph{sign test} for assessing, whether a
recommendation system $A$ significantly outperforms recommendation
system $B$, relative to some per user evaluation metric. For this
purpose, we count the number $n_A$ of \emph{users} where $A$ yields
better scores than $B$ and inversely $n_B$. We thereby ignore all
users with equal evaluation scores. The sign test assesses the
significance $p$ of an observed outcome $n_A$ and $n_B$ by estimating
the probability that $A$ is not truly better than $B$ with the
probability that $n_A$ out of $n\coloneqq n_A+n_B$ uniform Binomial
tries succeed:
\[
p = \left(\frac{1}{2}\right)^n\sum_{i=n_A}^n\frac{n!}{i!(n-i)!}
\]
Please refer to~\cite{shani2011evaluating,demsar2006statistical} for
further details and more advanced hypothesis tests for comparing
recommendation systems. Please note that we only assess the
significance of selected results in order to reduce the impact of
correcting the influence of multiple testings (\eg Bonferroni
correction).

\makeatletter{}\subsection{Usage Data}\label{sec:usage}
For our analysis and evaluation, we considered the \nameling's
activity log entries within the time range \emph{2012-03-06} until
\emph{2012-08-10}. In the following, we firstly describe the collected
usage data, analyze properties of emerging network structures and
finally compare interrelations between the different networks. 

In total, 38,404 users issued 342,979 search requests. Subsequently,
we differentiate between activities where a user manually entered a
given name (``\emph{Enter}''), followed a link to a name within a
result list (``\emph{Click}'') or added a given name the list of
favorite names (``\emph{Favorite}''). Table~\ref{tab:usage:basic}
summarizes high level statistics for these activity classes, showing,
\eg that $35,684$ users entered $16,498$ different given names.\comments{
\begin{table}
  \centering
  \caption{Basic statistics for the different activities within the \nameling.}
  \label{tab:usage:basic}
  \begin{tabular}{l|r|r}
                   &  \#Users &  \#Names \\\hline
    \enter         & $35,684$ & $16,498$ \\\hline
    \click         & $22,339$ & $10,028$ \\\hline
    \favorite      &  $1,396$ &  $1,558$
  \end{tabular}
\end{table}
}

\begin{figure*}
  \centering
  \includegraphics[width=0.32\linewidth]{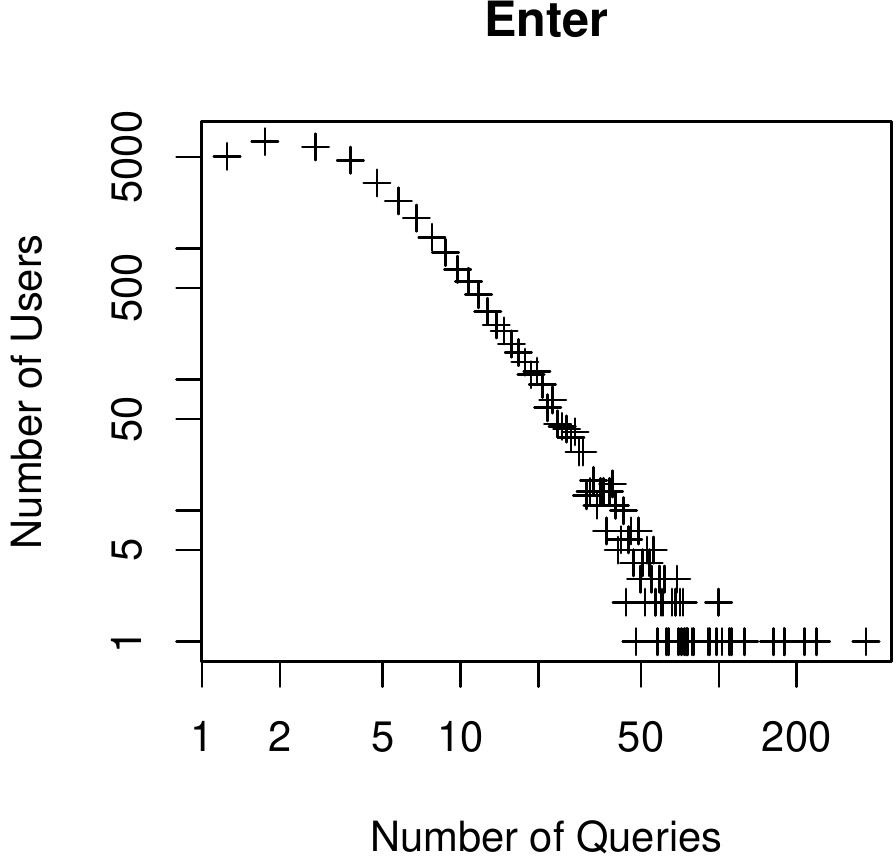}
  \includegraphics[width=0.32\linewidth]{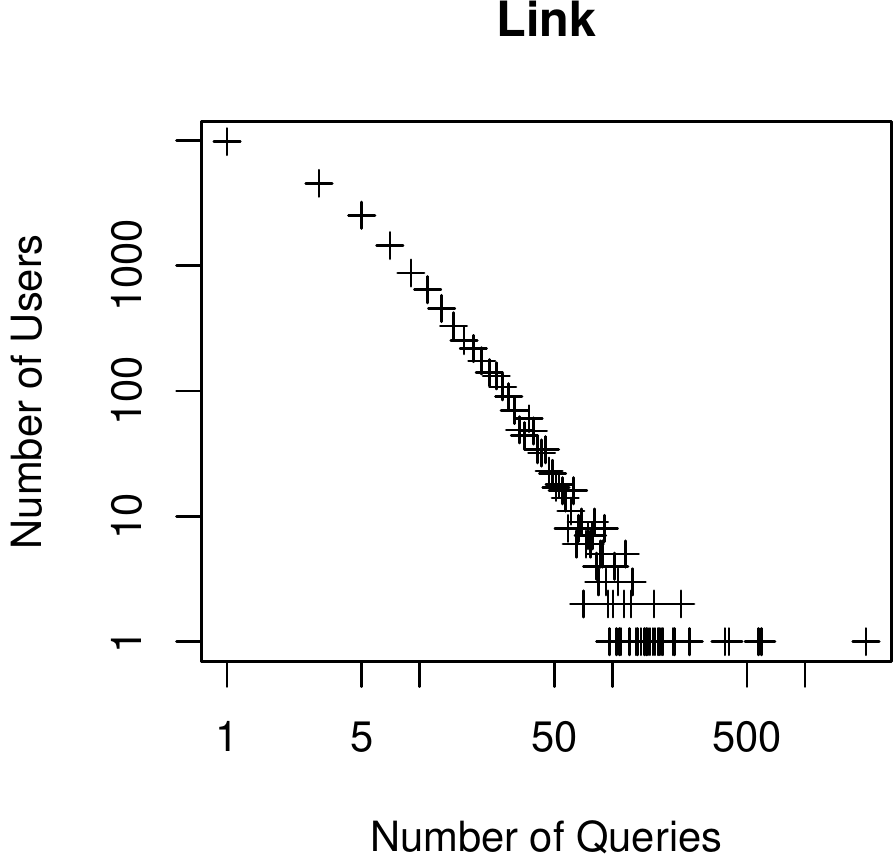}
  \includegraphics[width=0.32\linewidth]{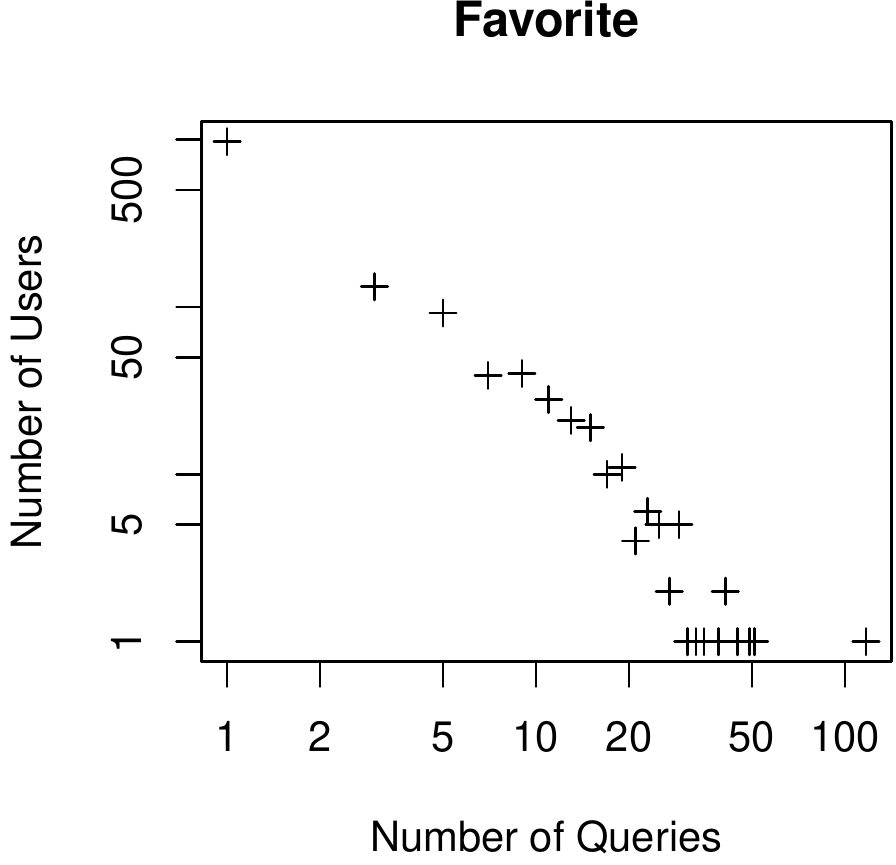}
  \caption{Number of users per query count for \enter, \click and \favorite activities.}
  \label{fig:usage:hist}
\end{figure*}

For analyzing how different users contribute to the \nameling's
activities, Figure~\ref{fig:usage:hist} shows the distribution of
activities over the set of users, separately for \enter, \click and
\favorite requests. Clearly, all activities' distributions exhibit
long tailed distributions, that is, most users entered less than 20
names but there are also users with more than 200 requests.
\enlargethispage*{2\baselineskip}
  
In order to assess the interdependence of name contexts established by
search queries within the \nameling and co-occurrences within
\wikipedia, we further constructed for each activity class
$C\in\set{\text{\enter},\text{\click},\text{\favorite}}$ a
corresponding weighted projection graph $G_{C}$, where an edge $(i,j)$
exists with weight $k$, if $k$ users searched for both names $i$ and
$j$). Table~\ref{tab:usage:networks} summarizes various high level
network statistics for all considered projection graphs. All
projection graphs encompass a giant connected
component~\cite{newman2003structure}, giving rise to a notion of
relatedness among names.
\begin{table}
  \centering
  \caption{High level statistics for all considered projection graphs
    with the number of weakly connected components \#wcc and
    largest weakly connected components lwcc.}
  \begin{tabular}{l|r|r|r|r|r}
             &  \multicolumn{1}{c|}{$|V|$}  
             &  \multicolumn{1}{c|}{$|E|$}         &   \#wcc  & lwcc & $D$ \\\hline\hline
    $G_{\text{\enter}}$
             &  $15,992$  & $996,930$       &     $72$ &   $15,834$ & $7$ \\\hline\hline
    $G_{\text{\click}}$
             &   $9,665$  & $2,633,220$     &     $79$ &    $9,465$  & $10$ \\\hline\hline
    $G_{\text{\favorite}}$
             &   $1,365$ &   $60,124$       &      $23$ &  $1,306$   & $7$   \\
  \end{tabular}
  \label{tab:usage:networks}
\end{table}
Please note that these usage graphs themselves can be used for
calculating similarity among given names, \eg by applying the same
similarity metrics as discussed for the co-occurrence
graphs. 

For assessing the interdependence of the name contexts established by
the different co-occurrence networks from Sec.~\ref{sec:networks} and
those emerging from \nameling's query logs, we apply again the
quadradic assignment procedure test (\cf
Sec.~\ref{sec:internetwork:qap}). Table \ref{tab:usage:qap} shows the
pairwise correlation scores for all considered networks.
\begin{table}
  \centering
  \caption{Graph level correlations for all pairs of considered
    networks. All observed correlations are significant according to
    the quadradic assignments procedure (QAP).}
  \begin{tabular}{l|r|r|r|r|r|r}
             & $G_{\text{\enter}}$
             & $G_{\text{\click}}$
             & $G_{\text{\favorite}}$
             & $\text{Wiki}^{DE}$
             & $\text{Wiki}^{EN}$
             & $\NTweets$
             \\\hline\hline
    $G_{\text{\enter}}$
             &   $-$     &    $0.466$    &    $0.370$    &    $0.138$    &    $0.054$   &     $0.0004$   \\\hline
    $G_{\text{\click}}$
             &           &     $-$       &    $0.364$    &    $0.110$    &    $0.056$   &     $0.0002$   \\\hline
    $G_{\text{\favorite}}$
             &           &               &      $-$      &    $0.040$    &    $0.032$   &     $0.0038$
  \end{tabular}
  \label{tab:usage:qap}
\end{table}
Again, correlations are more pronounced within a system (\eg $0.466$
for the $G_{\text{\enter}}$ and
$G_{\text{\click}}$). But the graph level correlations
for networks obtained from the \nameling exhibit significantly higher
correlation scores for the co-occurrence network obtained from the
German \wikipedia, indicating that the dominance of German users has
an impact on the emerging name contexts within the search based
networks which are more related to the accordingly language dependent
network structure within the co-occurrence network from
\wikipedia. Finally, the correlation scores for the \twitter based
co-occurrence networks shows the least magnitude, though still
signficant for $G_{\text{\enter}}$ and
$G_{\text{\favorite}}$ in terms of the \qap test
($p<0.05$). 

Summing up, we conclude that users accessing a name search engine like
\nameling are implicitly establishing name contexts which differ from
those, obtained from encyclopedic data sources, such as
\wikipedia. These user centric name contexts are more likely to
reflect the user's taste and personal preference and accordingly may
be used for generating personalized name recommendations.  
\makeatletter{}\subsection{The Recommendation Task}\label{sec:recommender}\enlargethispage*{3\baselineskip}
\comments{
In contrast to search engines, where the user queries for certain
keywords and expects in result a list of matching relevant items,
recommendation systems try to suggest suitable items of interest
automatically, \eg based previous activities or profile data of the
user.

Depending on the field of applications, the actual goal of a
recommendation system differs. In the context of folksonomies, \eg
\emph{tag recommendation systems} aim at reducing the cognitive effort
for assigning keywords (tags) to bookmarks, whereas \emph{product
  recommendation systems} suggest products to buy in an on-line
store. 

Accordingly, the assessment of what a good recommendation system 
is, strongly depends on the application context. Moreover, depending on
the point of view, different optimization criteria apply. Where the
on-line store, ultimately, may expect the recommendation system to
increase the annual income, the user might rather be interested in
cheaper product alternatives.

The present work tackles the task of recommending given names, aiming
at assisting future parents in finding and choosing a suitable given
name. In the following, we firstly introduce the considered
recommendation task and according objectives. We then discuss the
evaluation of according recommendation systems. 

\subsubsection{The Recommendation Task}\label{sec:recommender:task}
}
The choice of the given name is one of the first important decision
future parents have to make. Many influencing factors have an impact
on this decision process, such as, \eg cultural background, current
trends and personal taste. Typically, large collections of given names
with additional background information for the corresponding names are
at hand, either in the form of a lexicographical book or a specialized
web site.
Where a search engine for given names allows the user to browse
through the list of names ordered by some notion of relevance, a
recommendation system suggests a small personalized list of names
which the user might be interested in.

Different constraints can apply to the list of recommended names,
notably depending on the amount of background information
available. Given names from the user' ego-network of some on-line
social network, \eg could be ignored (assuming the user knows the given
names of her or his friends), or, a certain diversity of names
requested, as there is not desirable to present the user, \eg just different
variants of a single name.
Each of such constraint influences the assessment of the quality of a
recommendation system. Accordingly, there is no single valid
evaluation protocol. Below, we present the evaluation protocol we
applied for establishing baseline results for the quality of
recommendation systems for given names.

\paragraph{Evaluation Protocols}\label{sec:recommender:evaluation}
The evaluation of a recommendation system is strictly dependent on the
targeted objective~\cite{shani2011evaluating}. Ultimately, only an
appropriate comparative live evaluation of different recommendation
systems can comprehensively assess the performance of such systems. In
this section, we focus on the \emph{prediction accuracy} relative
to a publicly available activity log file from \nameling as presented
in Section~\ref{sec:usage}.

For this purpose, we process each user $u$ separately and remove from
the set of entered search names \enter{$(u)$} a subset \test[$(u)$]
for evaluation. We then use all activity data without the selected
evaluation data for calculating recommendations and assess the
prediction accuracy relative to
\test[$(u)$]. \comments{
\begin{figure}
  \centering
  \includegraphics[scale=0.3]{recommender/evaluation}
  \caption{The applied evaluation protocol. For user $u$ a subset
    \test[$(u)$] is selected for assessing the recommender's
    prediction accuracy.}
  \label{fig:recommender:evaluation:schema}
\end{figure}
}
We applied different strategies for selecting the evaluation set
\test[$(u)$], each of which suitable for examining certain research
questions:
\begin{description}
\item[\emph{TakeKIn}:]
  \test[$(u)$] is randomly sampled with size $\size{\enter{(u)}}-k$. 
\item[\emph{LeaveKOut}:]
  \test[$(u)$] is randomly sampled with size $k$.
\item[\emph{TakeFirstIn}:]
  \test[$(u)$] consists of the last $\size{\enter{(u)}}-k$ names the
  user $u$ has entered. Name recommendations are therefore based on
  the \emph{first} $k$ names.
\item[\emph{LeaveLastOut}:]
  \test[$(u)$] consists of the \emph{last} $k$ names the user $u$ has
  entered.
\end{description}
Whenever results among different number of known names $k\le N_{\max}$
are compared, we additionally removed $N_{\max}-k$ names from the
user's test set $\test{(u)}$ to ensure that results for different
values of $k$ are comparable and not determined through varying sizes
of $\test{(u)}$.

\comments{Subsequently, we focus on the \emph{TakeKIn} and
  \emph{TakeFirstIn} evaluation protocol, referring
  to~\cite{mitzlaff2013recommending} for a more detailed
  comparison. }

Please note that we only used the set of directly entered names
(\enter) for evaluation and did not use the combination of \enter and
\favorite events for evaluation, as these interrelations among names
have a strong bias induced by the ranking function which was
implemented in \nameling~\cite{mitzlaff2012namelings}.

\paragraph{Evaluation Metrics}\label{sec:recommender:evaluation:metrics}\enlargethispage*{2\baselineskip}
Various metrics for assessing the prediction accuracy of
recommendation systems exist and the choice of the applied metric
depends, among others, on the application context. Firstly, global
metrics like \map and \ndcg summarize the prediction performance of
all recommended items, whereas prefix metrics like \precision{@$k$}
and \ndcg{@$k$} only consider the first $k$ recommended
items.

For the present work, we differentiate the recommendation task from
searching by restricting recommendations to the corresponding top $k$
items, giving favor to \precision{@$k$} and \ndcg{@$k$}, whereof the
latter accounts for the ordering of the recommended items within the
top $k$ positions. For reference though, we also consider \map and
\ndcg. When comparing different recommendation systems based on the
corresponding \map scores, special care has to be taken if one of the
recommendation systems fails to generate rankings for all
items. Consider, \eg the \pagerank based \NR and user based
collaborative filtering \UCF. By construction, \NR assigns weights
to all names, whereas \UCF cannot infer weights for names which are not
connected to one of the queried user's search names. Now, even
for very low weights, relevant names are considered for calculating
the average precision score which may degrade \map significantly (due
to the sensitivity of average towards outliers). 

We argue that the fairest handling of such situations is to virtually
place all $\ell$ relevant names for a user $u$ which an recommendation
system $\text{Rec}$ failed to recommend, at the end of $\text{Rec}(u)$.

\todo{Other metrics exist 
  -> not all can be considered
  -> AUC: not directly applicable to PPR/PPR+ as for each user an
  independent ranking is calculated
}
 
\makeatletter{}\subsection{Experimentation}\label{sec:experiments}
This section focuses on the evaluation of the \emph{prediction
  accuracy} of different recommendation systems. On the one side, the
obtained results are a basis for deciding which recommendation system
should be used for recommending given names in a corresponding
application. On the other side, we thereby establish baseline results
for the task of recommending given names which serve as a reference
for developing and evaluating new approaches.

\subsection{Applied Recommender Systems}\label{sec:experiments:recommender}
Subsequently, we shortly summarize each considered recommendation
system and introduce according abbreviations which we henceforth use
for identifying individual recommendation systems.
\begin{description}
\item[User based CF (\UCF):] Adopting the \emph{weighted sum} approach
  for user based collaborative filtering~\cite{su2009survey} to the
  binary case (a user searched or searched not for a given name), we
  set
\[
\UCF(u,i)\coloneqq \sum_{v\in\mathcal{N}_i}\SIM(u,v).
\] 
For the \emph{nearest neighbor} approach, only the top $N$ similar
users to $u$ are considered in the summation.

For recommending $k$ names to user $u$, the top $k$ names, ordered
descending by $\UCF(u,*)$, are determined (ignoring names $j$ with
${r}_{uj}>0$).
\item[Item based CF (\ICF):]
  Adopting the \emph{weighted sum} approach
  in~\cite{sarwar2001itembased} to the binary case, we set:
\[
\ICF(u,i)\coloneqq
\frac{
  \sum_{j\in\mathcal{M}_u}\SIM(i,j)
}{
  \sum_{j\in\mathcal{M}}\SIM(i,j)
}.
\]
For recommending $k$ names to user $u$, the top $k$ names ordered
descending by $\ICF(u,*)$ are determined (ignoring names $j$ with
${r}_{uj}>0$).
\item[PPR/PPR+] The \emph{preferential \pagerank} similarity is based
  on the well known \pagerank~\cite{Brin98theanatomy} algorithm. For a
  $m\times m$ column stochastic adjacency matrix $A$ and damping
  factor $\alpha$, and uniform \emph{preference vector}
  $\vec{p}\coloneqq (1/m,\ldots,1/m)$, the \emph{global} \pagerank
  vector $\vec{w}\eqqcolon\PR$ is given as the fixpoint of the
  following equation:
  \[
  \vec{w} = \alpha A\vec{w} + (1-\alpha)\vec{p}
  \]
  In case of the \emph{preferential \pagerank} for a given set of
  nodes $\mathcal{I}$, only the corresponding components of the
  preference vector are set and we set accordingly $\PPR(\mathcal{I})$
  to the fixpoint of the above equation with
  \[
  p_i \coloneqq
  \begin{cases}
    \frac{1}{\size{\mathcal{I}}},&\text{if } i\in\mathcal{I}\\
    0, & \text{otherwise.}
  \end{cases}
  \]
  As a new item based recommendation approach we propose \NR (for
  brevity abbreviated with \PPRa), which is an adoption of the idea
  presented in~\cite{hotho2006information}, where the global \pagerank
  score $\PR$ is subtracted from the preferential \pagerank score in
  order to reduce frequency effects and
  set \[\PPRa(\mathcal{I})\coloneqq\PPR(\mathcal{I})-\PR.\] For
  recommending $k$ names to user $u$, we calculate
  $\vec{w}\coloneqq\PPRa(\mathcal{M}_u)$ on the column stochastic
  adjacency matrix derived from $G_{\enter}$ (\cf Sec.~\ref{sec:data})
  and recommend the top $k$ names ordered descending by $\vec{w}$,
  thereby ignoring names $j$ with ${r}_{uj}>0$.

  Please note that we show results obtained by averaging the rankings
  from individual query names, \ie
  \[\frac{1}{\size{\mathcal{I}}}\sum_{j\in\mathcal{I}}\PPRa(\set{j})\]
    which showed the same results as $\PPRa(\mathcal{I})$. 
\item[WRMF] The \emph{weighted matrix factorization
  method}~\cite{hu2008collaborative} is designed to deal with implicit
  feedback scenarios like the our evaluation setting described
  in~\ref{sec:recommender:evaluation}. The observed user-item
  interactions $r_{ui}$ are interpreted as indicators of user $u$'s
  preference for item $i$ which is associated with a certain level of
  confidence, depending on the amount of observed
  interactions. Unobserved interactions $\hat{r}_{ui}$ are predicted
  by building a factor model of the user-item matrix $R$ via an
  regularized learning method with alternating least squares
  optimization. 
\item[MostPopular]
  The \mostpopular recommendation approach predicts for all users the
  most popular names, \ie the top $k$ names ordered decreasing by
  frequency.
\item[Random]
  The Random recommendation approach randomly selects $k$ names which
  the user has not searched before.
\end{description}
The item-based and user-based collaborative filtering as well as the
\mostpopular approaches were implemented using the machine learning
library \mahout\footnote{\url{http://mahout.apache.org}}. We evaluated
various similarity metrics, showing here results for the
Log-Likelihood similarity~\cite{dunning1993accurate} which
outperformed the others. The \pagerank-based approaches \PPR and \PPRa
are implemented using the
\eigen~\footnote{\url{http://eigen.tuxfamily.org}} library for matrix
operations. For WRMF we applied the implementation of the
\mymedia~\cite{gantner2011mymedialite} library with the corresponding
default parametrization.

We also considered the Bayesian personalized ranking approach for
matrix factorization~\cite{rendle2009bpr}, additionally with soft-margin
optimization as proposed in~\cite{weimer2008improving} as implemented
in the \mymedia recommendation library. But the obtained result showed
worse prediction performance than the simple \mostpopular
recommendation approach and are therefore excluded from the
presentation below for clarity.
\paragraph{Evaluation Data}\label{sec:experiments:data}
For evaluating the prediction accuracy of the considered
recommendation systems, we applied the different evaluation protocols
described in Section~\ref{sec:recommender:evaluation} for
$k=1,\ldots,N_{\max}$ with $N_{\max}\coloneqq 10$. To assure that for
every user $u$ there are at least five names available for testing and
prediction, respectively, we restricted our evaluation to users who
entered more than 14 different names (that is, $\size{\enter{(u)}}\ge
15$), leaving 1,230 users for evaluation. Please note that we neither
considered the set \click{$(u)$} of names which the user clicked on
nor the set \favorite{$(u)$} which the user added to the list of
favorite names, as these data sets are strongly biased by the ranking
which was implemented in \nameling's search back-end.

\subsection{Results}\label{sec:recommender:experiments:results}
We now present our results on the prediction accuracy of the
considered recommendation systems. We firstly present the results for
the TakeKIn and LeaveKOut experiments where the results are obtained
relatively to the \emph{number} of names and finally the results of
the TakeFirstIn and LeaveLastOut experiments, where results are
obtained relative to the chronologically ordered \emph{first} names of
a user's search history.

\paragraph{TakeKIn}\label{sec:experiments:results:takekin}
This experiment aims at comparing the impact of personalization on the
prediction accuracy of the different recommendation systems. In
particular we want to examine...
\begin{compactitem}
\item ...whether and to which extend the performance of a
  recommendation system benefits from an increasing number of known
  names (\ie increasing $k$).
\item ...whether for new users with only few known names, different
  recommendation systems perform best than for those users with a
  large search history.
\end{compactitem}
For this purpose, we applied the TakeKIn evaluation protocol as
described in Section~\ref{sec:recommender:evaluation} For eliminating
the influence of certain selections of known names, we repeatedly
sampled (25 times) for each user the set of $k$ known names and
averaged the resulting evaluation scores. Furthermore, we removed
$N_{\max}-k$ names from the user's test set $\test{(u)}$ to ensure
that results for different number of known names $k$ are comparable
and not determined through varying sizes of $\test{(u)}$ for different
$k$.

\begin{figure*}
  \centering
  \subfloat[][\precision{@$k$}]{\label{fig:experiments:takekin:p5}
    \includegraphics[scale=0.55]{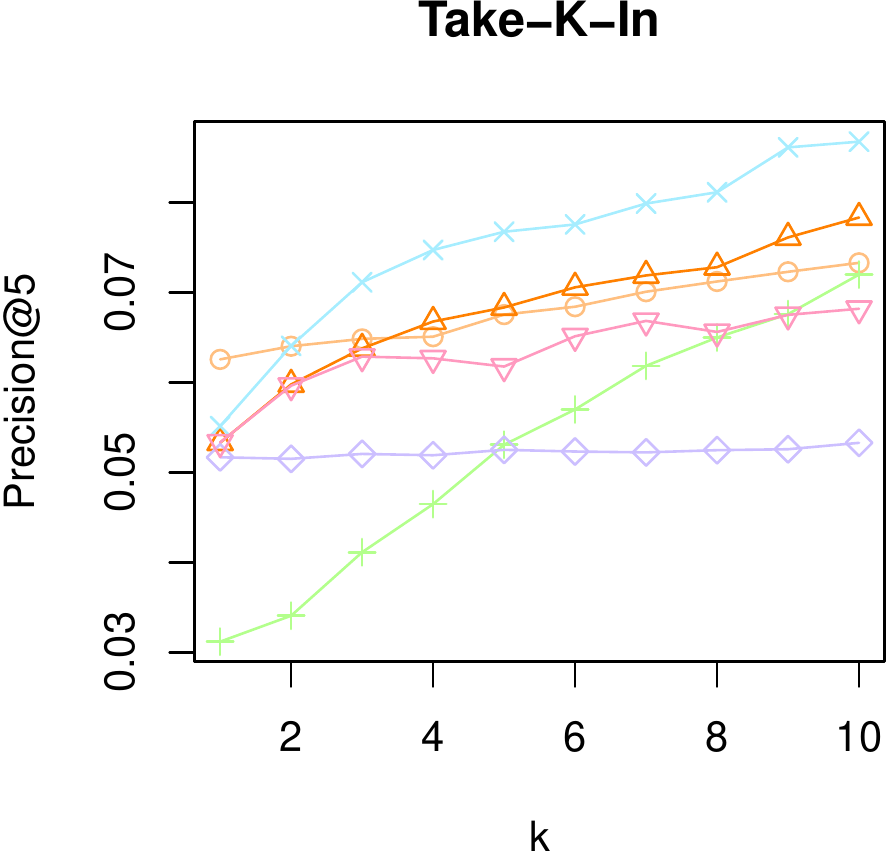}
  }
\comments{
  \subfloat[][\recall{@$k$}]{\label{fig:experiments:takekin:r5}
    \includegraphics[scale=0.55]{takeKIn-rec5-crop}
  }
}
  \subfloat[][\ndcg{@$k$}]{\label{fig:experiments:takekin:ndcg5}
    \includegraphics[scale=0.55]{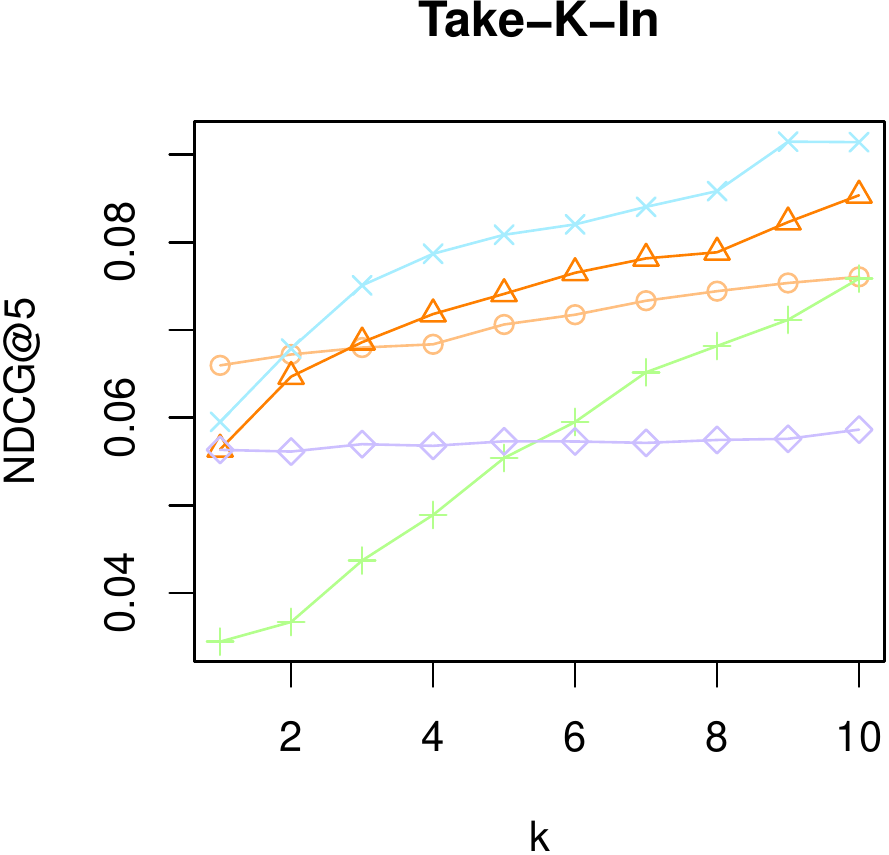}
  }
  \subfloat[][\map]{\label{fig:experiments:takekin:map}
    \includegraphics[scale=0.55]{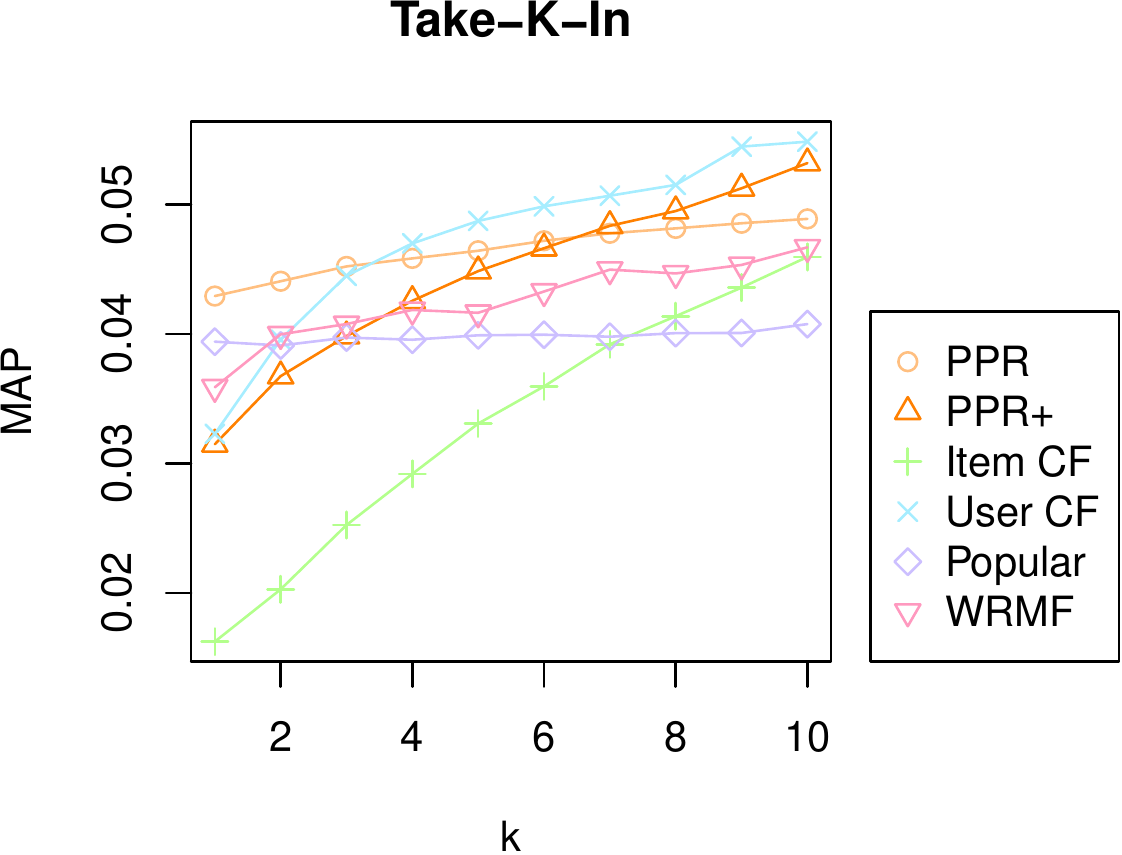}
  }
  \caption{Precision accuracy of the different recommendation systems,
    relative to varying number of known names per user. For clarity of
    presentation the random baseline was omitted which lies strictly
    below $10^{-3}$ for all metrics.
  }
  \label{fig:experiments:takeKIn}
\end{figure*}

Figure~\ref{fig:experiments:takeKIn} shows the obtained results for
\precision{@5}, \ndcg{@5} and \map. Please note that we didn't include the
random baseline results into Figure~\ref{fig:experiments:takeKIn} for
clarity, as all corresponding evaluation scores were
two magnitudes below the worst result of the other recommenders.

We begin with a discussion of some general observations. Firstly, all
obtained performance scores are low in magnitude, indicating that
given names are hard to predict based on the user's search
history. One of the key difficulties in such \emph{implicit feedback
  data} is due to the indistinguishable intent of search
requests~\cite{hu2008collaborative}: A user might search for a name
because the user likes the name, or the user might search for names he
doesn't like and just wants to explore the names
neighborhood. Equally, the fact that a user did not search for a
certain name might either express that the user dislikes the name, was
not aware of the name or just did not search for the given name until
now but eventually will.

\begin{figure}
  \centering
  \includegraphics[scale=0.45]{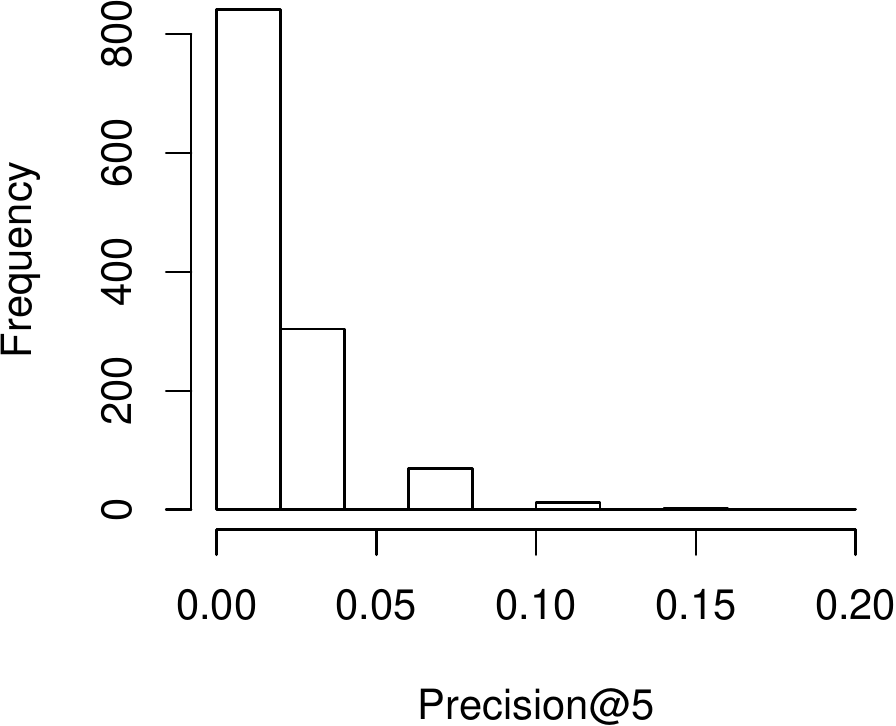}
  \includegraphics[scale=0.45]{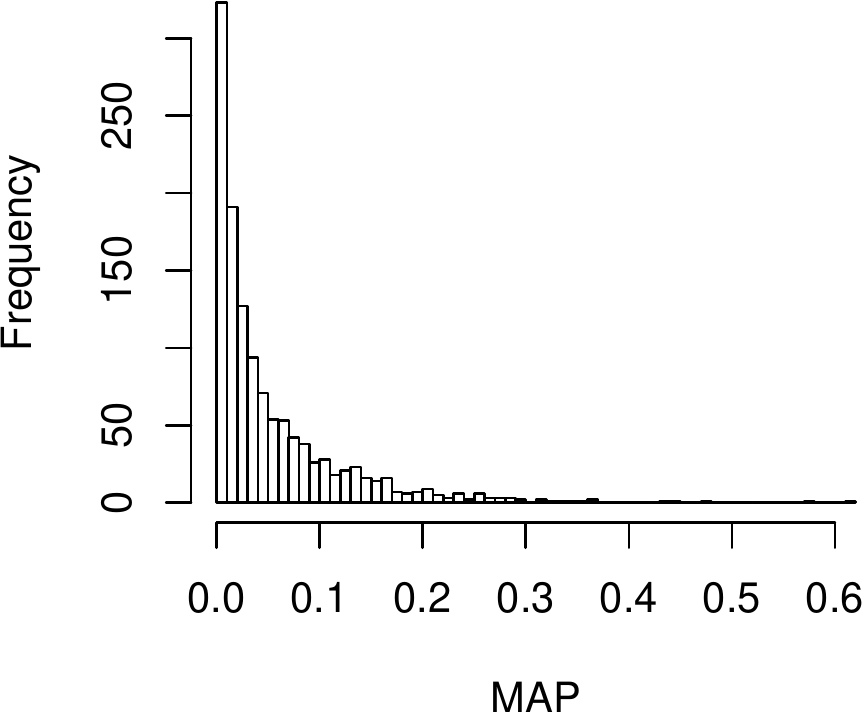}
  \caption{ 
    Distribution of the \precision{@5} and \map scores for \PPRa with ten known names.  }
  \label{fig:experiments:hist}
\end{figure}

Figure~\ref{fig:experiments:hist}
exemplarily presents the distribution of the \precision{@5} and \map
scores for \PPRa with ten known names, showing that the low scores are
due to a highly skewed distribution where for most users no or only
few relevant names are predicted. This, on the other hand, emphasizes
the importance of additional analysis beside calculating average
prediction metrics, as the average over skewed distributed values is
sensitive to outliers. Nevertheless, the considered performance
metrics consistently assess the impact of an increasing number of
known names for all considered recommendation systems.

For a more formal analysis of the consistency among the considered
performance metrics, we assessed the prediction accuracy of each
considered recommendation setting (in total 1435 models) with every
performance metric, respectively averaged over all users. We thus
obtained for each metric a ranking on all models, among which we
calculated Kendall's $\tau$ rank correlation
coefficient. Table~\ref{tab:experiments:correlation:models} shows the
calculated correlation scores. Notably, all metrics but \ndcg{@10}
show a very high correlation, supporting the observed consistent
assessment of different recommendation systems with varying number of
known names in Figure~\ref{fig:experiments:takeKIn}.

\begin{table*}\small
  \centering
  \caption{Correlations between the different evaluation metrics
    averaged for each model separately.}
  \makeatletter{}\begin{tabular}{rrrrrrrr}
  \hline
               & Precision@5 & Precision@10 & Recall@5 & Recall@10 & NDCG@5 & NDCG@10 & MAP \\ 
  \hline
  Precision@5  & -           &               &          &           &        &         &     \\ 
  Precision@10 & 0.94        & -             &          &           &        &         &     \\ 
  Recall@5     & 0.89        & 0.87          & -        &           &        &         &     \\ 
  Recall@10    & 0.85        & 0.88          & 0.94     & -         &        &         &     \\
  NDCG@5       & 0.94        & 0.90          & 0.86     & 0.82      & -      &         &     \\
  NDCG@10      & 0.37        & 0.38          & 0.34     & 0.34      & 0.36   & -       &     \\
  MAP          & 0.79        & 0.79          & 0.79     & 0.79      & 0.81   & 0.35    & -   \\
  \hline
\end{tabular}
 
  \label{tab:experiments:correlation:models}
\end{table*}

We now turn our focus to the discussion of the performance scores for
the different recommendation systems in
Figure~\ref{fig:experiments:takeKIn}. The \emph{most popular}
recommendation approach shows a relative good performance, as expected
independent of the number $k$ of known names. This is therefore a
suitable baseline for assessing other recommendation systems'
performance scores. All other considered methods benefit from
increasing $k$. In particular the \emph{item based
  collaborative filtering} shows a linearly increasing performance
scores, thereby showing worse performance than the most popular
baseline for lower $k$. Except for $k\le2$, the \emph{user based
  collaborative filtering} approach shows the highest performance
scores. 

As for \PPR and \PPRa, the former shows better performance for lower
$k$ (even outperforming user based collaborative filtering) whereas
the latter steadily increases with the number of known names. Please
note that this is in line with the intention for the construction of
\PPRa: the plain preferential \pagerank is strongly influenced by
global frequencies which are reduced by subtracting the global
\pagerank (\cf Section~\ref{sec:experiments:recommender}). As the
popular (\ie most frequent names) are already a relative good
recommendation, the influence of global frequencies is desirable if
only few names of a user are known.

Finally WRMF also shows increasing prediction accuracy with
increasing number known names.

\paragraph{LeaveKOut}\label{sec:experiments:results:leavekout}
In this experimental setup, as much of the user $u$'s search history as
possible is used for predicting the user's evaluation data
\test{$(u)$}. We may therefore derive from this experiment, which
recommendation system yields (in average) the best prediction accuracy
in a running system where users have broadly varying search history
sizes.

Table~\ref{tab:experiments:results:leavekout} shows the performance
scores for predicting ten randomly chosen names, based on the
remaining search log. We apply the sign test (\cf
Section~\ref{sec:preliminaries}) using the \map scores to test the
stated observations for significance and provide the correspondingly
obtained $p$-values. Firstly we note that the trends observed in the
previous experiment are affirmed. In particular \PPRa outperforms all
but the user base collaborative filtering ($p<10^{-3}$). In case of
user based collaborative filtering, the average performance scores
indicate the \PPRa yields better recommendations. But considering the
per user performance, \ie the number of users where \PPRa yields
better results, the sign test indicates that \UCF performs best
($p<10^{-3}$).

\begin{table}\small
  \centering
  \caption{Leave-$10$-Out evaluation for all considered recommendation systems.}
  \makeatletter{}\begin{tabular}{rrrrr}
  \hline
 & Precision@5 & Recall@5 & NDCG@5 & MAP \\ 
  \hline
PPR & 0.065 & 0.039 & 0.067 & 0.045 \\ 
  PPR+ & \textbf{0.076} & 0.044 & \textbf{0.084} & \textbf{0.052} \\ 
  User CF & 0.074 & \textbf{0.045} & 0.079 & 0.050 \\ 
  Item CF & 0.062 & 0.034 & 0.066 & 0.042 \\ 
  MostPopular & 0.046 & 0.027 & 0.050 & 0.037 \\ 
  WRMF & 0.058 & 0.035 &  & 0.041 \\ 
  Random & 0.000 & 0.000 & 0.000 & 0.001 \\ 
   \hline
\end{tabular}
 
  \label{tab:experiments:results:leavekout}
\end{table}

\comments{
\begin{table*}\small
  \centering
  \makeatletter{}\begin{tabular}{rrrrrrrr}
  \hline
               & Precision@5 & Precision@10 & Recall@5 & Recall@10 & NDCG@5 & NDCG@10 & MAP \\ 
  \hline
  Precision@5  & -           &               &          &           &        &         &     \\ 
  Precision@10 & 0.78        & -             &          &           &        &         &     \\ 
  Recall@5     & 0.88        & 0.87          & -        &           &        &         &     \\ 
  Recall@10    & 0.70        & 0.72          & 0.78     & -         &        &         &     \\
  NDCG@5       & 0.94        & 0.75          & 0.86     & 0.68      & -      &         &     \\
  NDCG@10      & 0.83        & 0.91          & 0.76     & 0.80      & 0.84   & -       &     \\
  MAP          & 0.68        & 0.73          & 0.63     & 0.65      & 0.68   & 0.75    & -   \\
  \hline
\end{tabular}
 
  \caption{Correlations of the different evaluation metrics for all
    users in the Leave-$10$-Out evaluation setting.}
  \label{tab:experiments:results:leavekout:correlations}
\end{table*}
}

\paragraph{TakeFirstIn/LeaveLastOut}\label{sec:experiments:results:history}\enlargethispage*{1\baselineskip}
The previous experiments randomly selected names from a user's
search history for predicting the user's remaining names. This
approach has two advantages: Firstly, the effect of ordering of the
entered names is averaged out and the discussion is therefore focused
on the influence of merely the \emph{size} of set of known
names. Secondly, the repeated randomization smooths the results, thus
reducing the effect of the relative small population size (in the
considered setup only 1,230 users).

Nevertheless, the order of input is the one which a recommendation
system within a live setting has to deal with. We therefore also
consider the TakeFirstIn/LeaveLastOut evaluation protocol where the
chronological order of the search history is used for splitting the
evaluation data (\cf Section~\ref{sec:recommender:evaluation} for more
details).

\begin{figure*}
  \centering
  \subfloat[][\precision{@$k$}]{\label{fig:experiments:takefirstin:p5}
    \includegraphics[scale=0.55]{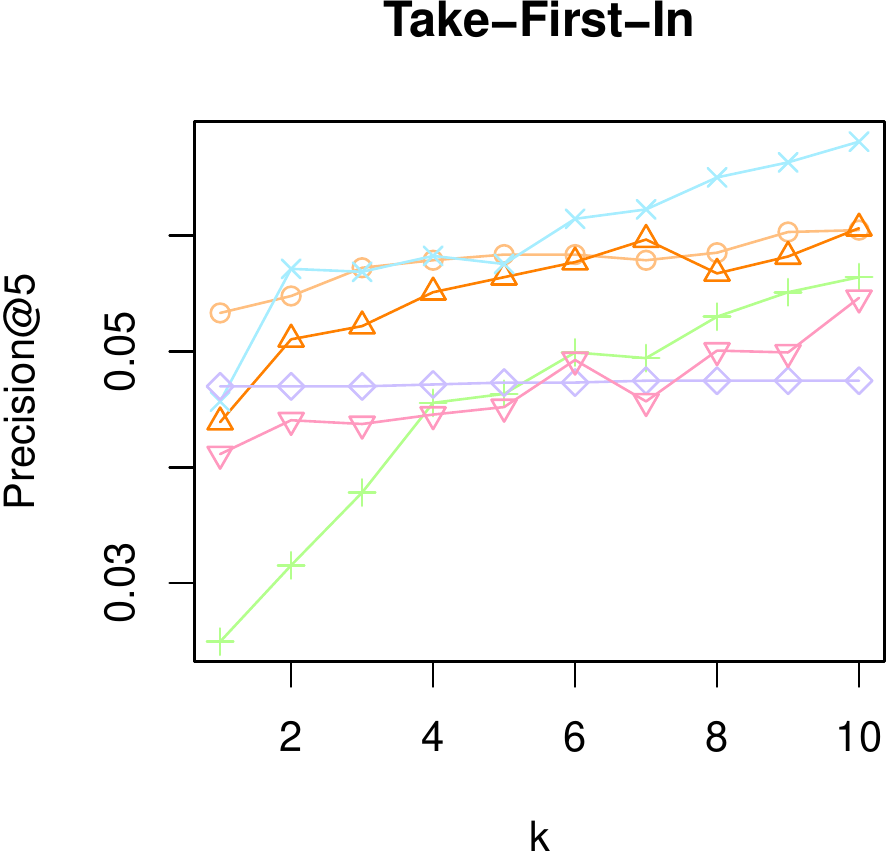}
  }
\comments{
  \subfloat[][\recall{@$k$}]{\label{fig:experiments:takefirstin:r5}
    \includegraphics[scale=0.55]{figs/eval/takeFirstIn-rec5-crop}
  }
}
  \subfloat[][\ndcg{@$k$}]{\label{fig:experiments:takefirstin:ndcg5}
    \includegraphics[scale=0.55]{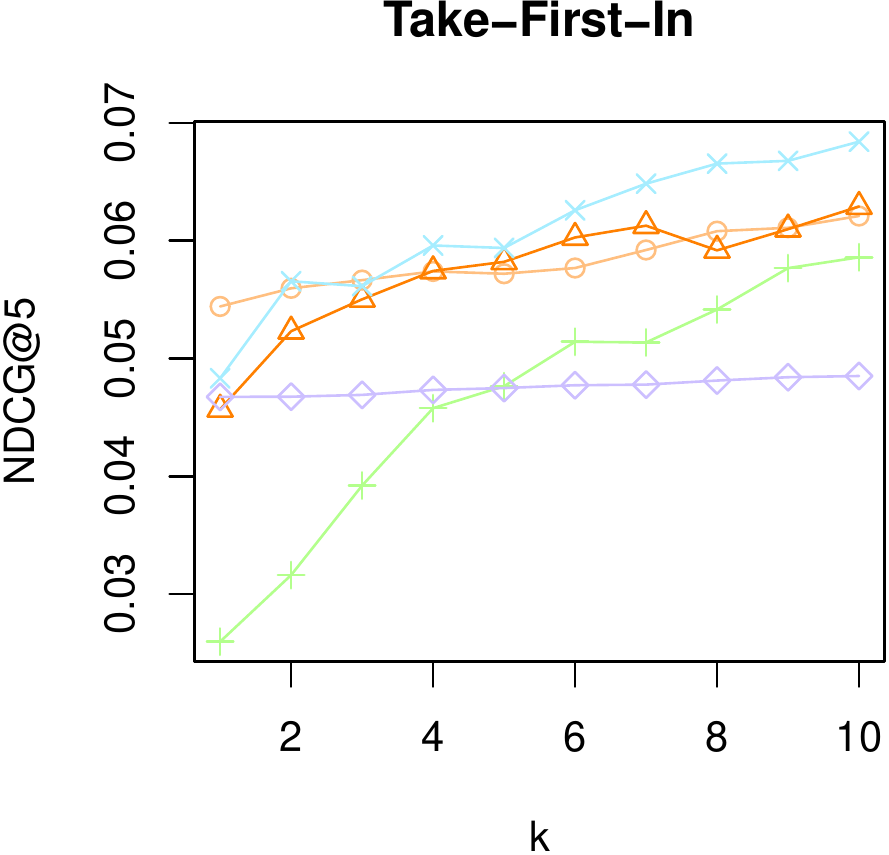}
  }
  \subfloat[][\map]{\label{fig:experiments:takefirstin:map}
    \includegraphics[scale=0.55]{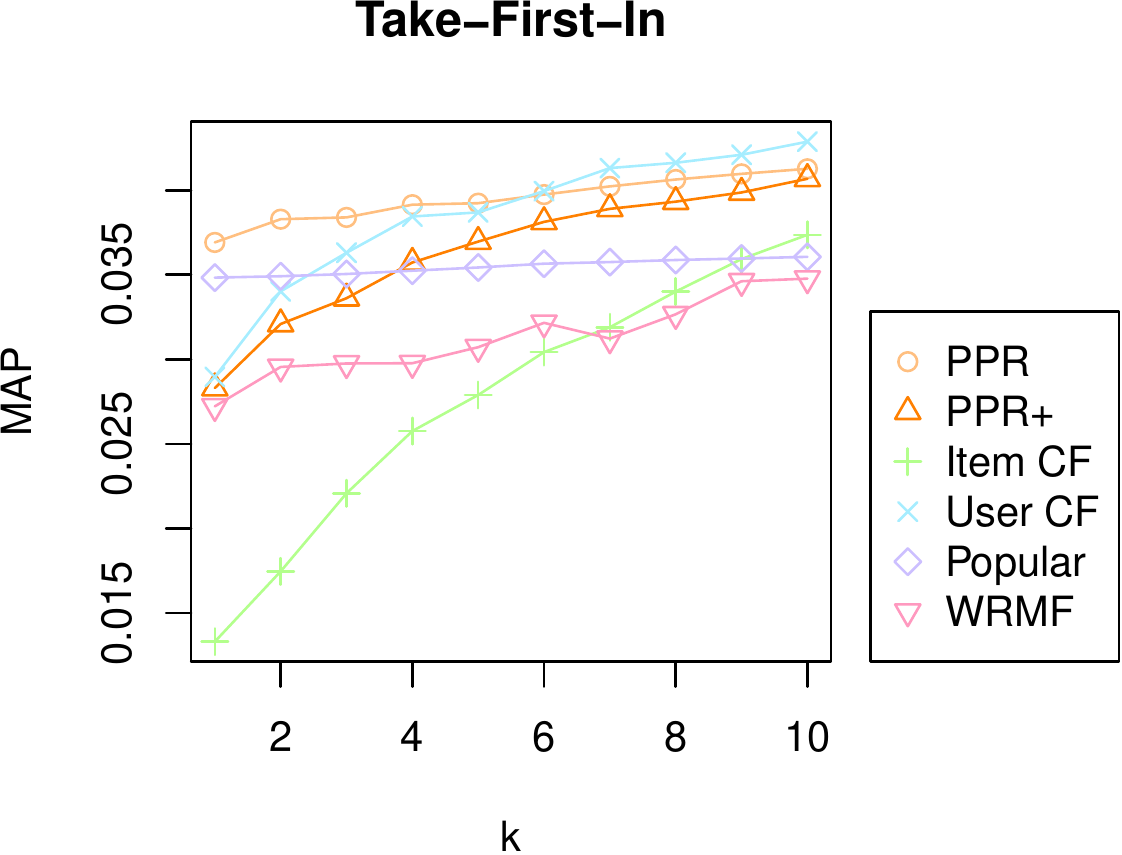}
  }
  \caption{Precision accuracy of the different recommendation systems,
    relative to varying number of known names per user.}
  \label{fig:experiments:takefirstin}
\end{figure*}

\begin{figure}
  \centering
  \includegraphics[scale=0.55]{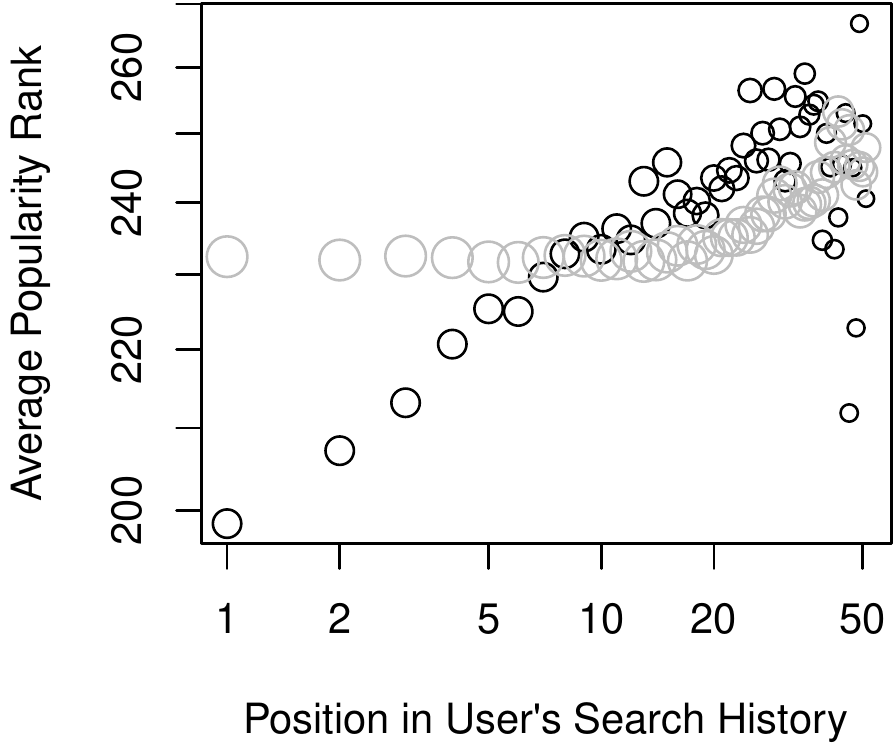}
  \caption{Average \emph{popularity rank} for all entered names
    relative to the corresponding position within the users' search
    history, indicating that users tend to search first for popular
    names. To rule out statistical effects, we shuffled the user's
    search history and, depicting the corresponding results in Grey.}
  \label{fig:experiments:popularityplot}
\end{figure}

Figure~\ref{fig:experiments:takefirstin} shows \precision{@5},
\ndcg{@5} and \map for the TakeFirstIn experiment.
Firstly, the result plots are less smooth than the corresponding plots
for the TakeKIn experiments as the additional smoothing induced by
repeated randomization is missing. Secondly, the relative assessment
of \UCF, \ICF, \MP, \RND and \WRMF is consistent with those of the
randomized TakeKIn experimentation. Most interestingly, \PPR performs
better for all $k$ than $\PPRa$ (in contrast to the results of the
TakeKIn experimentation). This, on the one hand, shows, that the order
of input does indeed matter. The worse assessment of \PPRa in the
chronological order hints at the impact of global frequencies (\ie
popularity of names), as by construction, \PPRa increases
personalization by reducing the influence of global frequencies. The
results in Figure~\ref{fig:experiments:takefirstin} indicate, that the
first entered names tend to be more popular names and later a name is
entered, the more specific the name is for the user. To underpin this
assumption, we calculated for each name its \emph{popularity rank},
that is, its position within the list of names which is decreasingly
ordered by search frequency. We than calculated for each chronological
position $p$ the average popularity rank of all names which were
search by some user at position $p$ within the user's search
history. Figure~\ref{fig:experiments:popularityplot} shows a positive
correlation between the position in the user's search history and the
popularity rank of the corresponding name, indeed indicating that
users tend to search first for popular names.

For reference we also summarize the prediction accuracy of the
different recommendation systems for the LeaveLastOut protocol, which
uses as much of the chronologically ordered names from the users
search history for predicting the remaining (last) names in
Table~\ref{tab:experiments:results:leavelastout}. The indicated
performance scores support observed the tendencies in the preceding
TakeFirstIn experiment.
\begin{table}\small
  \centering
  \caption{Leave-Last-Out evaluation for all considered recommendation systems.}
  \makeatletter{}\begin{tabular}{rrrrr}
  \hline
 & Precision@5 & Recall@5 & NDCG@5 & MAP \\ 
  \hline
PPR & 0.050 & 0.031 & 0.049 & 0.037 \\ 
  PPR+ & 0.052 & 0.031 & 0.053 & \textbf{0.038} \\ 
  User CF & \textbf{0.056} & \textbf{0.034} & \textbf{0.055} & \textbf{0.038} \\ 
  Item CF & 0.049 & 0.027 & 0.051 & 0.034 \\ 
  MostPopular & 0.041 & 0.025 & 0.041 & 0.032 \\ 
  WRMF & 0.047 & 0.028 &  & 0.032 \\ 
  Random & 0.000 & 0.000 & 0.000 & 0.001 \\ 
   \hline
\end{tabular}
 
  \label{tab:experiments:results:leavelastout}
\end{table}

\begin{figure}
  \centering
  \includegraphics[scale=0.55]{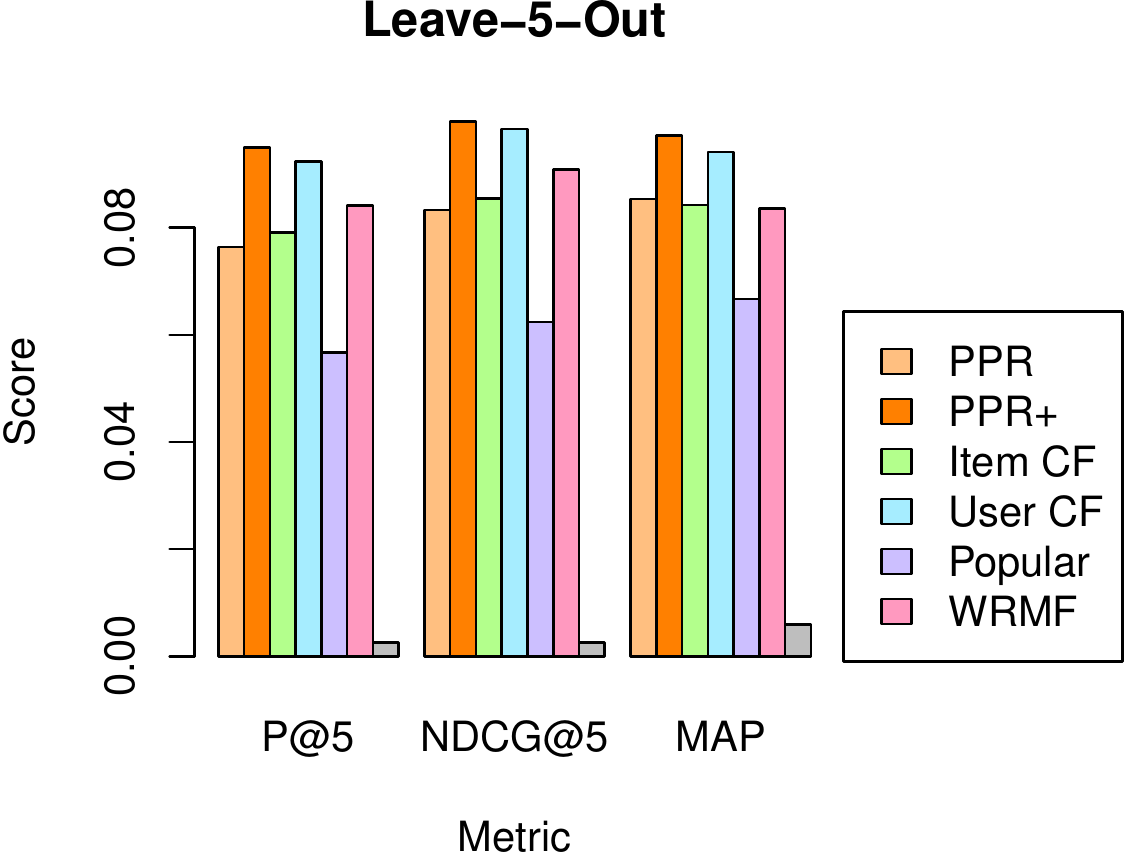}
  \caption{Average performance metrics for the favorite names
    (Leave-5-Out with 25 repetitions and minimal number of favorite
    names per user is 8).}
  \label{fig:experiments:favorites}
\end{figure}

For further supporting the observed characterizations of the different
recommendation systems with respect to prediction accuracy, we applied
the same evaluation protocol to the set $\favorite(u)$ names per user
$u$. As the corresponding evaluation data is much more sparse then the
set \enter(u) of entered names, we only show the results for the
LeaveKOut protocol with $k=5$ and considering only users with at least
eight favorite names (resulting in 230
cases). Figure~\ref{fig:experiments:favorites} shows the corresponding
results for \precision{@5}, \ndcg{@5} and \map for all considered
recommendation system respectively. The overall relative assessment of
the different system is consistent with the results obtained on the
set of entered names - only that \PPRa is now even outperforming \UCF
significantly ($p=0.0362$).

\subsection{Discussion}\label{sec:experiments:discussion}
In this section we presented results of an experimental setup which
aims at assessing the prediction accuracy of different recommendation
systems. Beside establishing baseline results for reference, the
presented evaluation guides the decision process of choosing a
recommendation system for integration in a running application like
\nameling. 

Considering just the prediction accuracy, the standard preferential
\pagerank consistently shows best performance for users for which only
one or two names are known. In any other case, the user based
collaborative filtering approach outperforms all other systems. As it
is important to generate good recommendations as soon as possible to
catch the user's interest, it is worth to combine these approaches,
recommending popular names to new users, \PPR based recommendations
for users with a small search history and \UCF otherwise.

But beside prediction accuracy, among others, run time efficiency is
an important issue for recommendation systems. In the context of tag
recommendation systems it was observed that, in a live setting, many
systems fail to provide recommendations within a justifiable time
limit~\cite{jaeschke2009testing}. Also the handling of new items (\ie
names) and users has to be taken into account. In the case of name
recommendations, the set of names is nearly constant over time, while
there is a steady stream of new users using the system while there are
very few users using the system for a long period of time. This is due
to the fact that the need for a given name typically ends after a
short time limit.

Considering the run time efficiency, \WRMF is a candidate due to the
constant time operations with the latent user and item vectors. But
also \PPR an \PPRa are efficient to implement, as the results are just
averaged over the individual query names of the user's search
history. This operation can already be implemented within a database
system which only holds the top similar names $\PPRa(i)$ and
$\PPR(i)$, respectively, for every name $i$. As the set of names is
nearly constant in time, good recommendations for new users can be
produced on these precomputed vectors.

Considering the results obtained on the chronologically ordered search
history of users, $\PPR$ yields results as least as good as $\PPRa$,
as depicted in Figure~\ref{fig:experiments:takefirstin}. But the
results obtained on randomized samples of the user's search history
show significant better performance of $\PPRa$. The observations in
Section~\ref{sec:experiments:results:history} indicate that the
difference in the prediction accuracy of \PPR and \PPRa between the
chronological and randomized evaluation can be explained with the
tendency that users tend to search first for popular names and later
more special names. But this search habits might change with the
availability of good recommendation systems. An important reason
for the popularity correlated search order is due to the fact, that
popular names are more known and a user first thinks of popular names
while browsing the system for more suitable names. A good
recommendation system will help the user in accessing suitable (\ie
personalized) names more directly.

Summing up, our adaption \PPRa of the preferential \pagerank is the
most promising candidate for implementation in a running
application. The choice of \PPRa is even more supported by considering
the actual favorite names in \nameling, where it even significantly
outperforms \UCF ($p=0.04$).

\comments{
\todo{Hints at: TFIDF-like optimizations for identifying
  characteristic names for users OR chonologically later}
\todo{PPR better PPRa hints at: popular names entered first 
  experiment
DISCUSS:
  ie: the names the user KNOWS
  here: recommender may heavily influence search behavior -> LIVE EVALUATION
}
\todo{discussion: popular names certainly important!}
}

\comments{
  MergedModel-Random:
  ## P@5:
  ## Avg   : 0.1406336
  ## Winner: 243 370 262 328 207 337

179.00000000 358.00000000 236.00000000 261.00000000 191.00000000
[6]   5.00000000   0.06719422

  MergedModel-History:
  ## P@5::
  ## avg: 0.121544
  ## win: 762 771 801 761 726 786

}

            \makeatletter{}\section{Diversification of PageRank based Recommendations}\label{sec:diversification}

\makeatletter{}The apparent diversity-accuracy dilemma of recommender systems is
especially relevant for the task of recommending names, as future
parents are often interested in names they like, but which should not
be too common within their environment. Concerning prediction
accuracy, the simple recommender which constantly recommends only the
popular names, already yields reasonable results~(\cf
\ref{sec:recommender:experiments:results}), but is useless for the
parent’s need to discover names which are suitable for their children
but are yet unknown to them.

So far, we only considered the \emph{prediction accuracy} for
assessing the quality of the different recommendation approaches. This
section aims at comparing the recommendation systems with respect to
the \emph{diversity} of the obtained recommendations. For this, we
firstly fix our notion of diversity and corresponding evaluation
measures and subsequently use these measures for comparing the
diversity of the most promising recommendation approaches from
Section~\ref{sec:recommender:experiments:results}.

Finally, we consider different ways for increasing diversity of our
\nameRank approach by incorporating background information based on
networks of given names obtained from \wikipedia, \twitter and
\nameling. 

\makeatletter{}\subsection{Asessment of Diversity}
Diversity of a set of recommended items can either be assessed
relative to \emph{semantic properties} (such as gender or cultural
context in the case of given names) or relative to its
\emph{adaptation} to the user's personal taste (\ie how diverse 
recommendations for different users are).

We focus on the aspect of \emph{adaptation}, because we distinguish
between the task of \emph{ranking} and \emph{recommendation} in the
context of our work, at which the former targets the use case of
searching and browsing for given names and accordingly benefits in
particular from semantical result diversification, whereas the latter
targets the use case of providing a small set of suitable names which
adopts primarily to the user's personal taste.

Assuming diversity of the user's personal taste, personalization of a
given recommendation system \rec can be assessed by measuring the
inter-user diversity for different users $u$ and $v$. Accordingly, we
consider the \emph{personalization index} $h$, proposed
in~\cite{zhou2010solving}
$
h(u,v)\coloneqq 1 - \frac{\reco{k}{u}\cap \reco{k}{v}}{k}.  $, where
$\reco{k}{u}$ denotes $\rec$'s top $k$ recommended items for user $u$.
Identical results for $u$ and $v$ thus yield a value of $h(u,v)=0$
whereas completely different recommendations yield a value of
$h(u,v)=1$. The overall personalization performance $h(\rec)$ is
assessed by averaging over all pairs of users.

\comments{
Additionally we consider the ability of a recommendation system to
generate unexpected results, \ie to recommend names which are (yet)
not well known. We assess the ``\emph{surprisal}'' of recommendation
results by means of the self-information $I(j)$ of an item $j$ as
introduced in~\cite{zhou2010solving}: $ I(j) \coloneqq
\log_2\left(\frac{\size{\mathcal{N}}}{\size{\mathcal{N}_j}}\right) $,
where $\mathcal{N}$ denotes the set of users and $\mathcal{N}_j$ the
set of users who expressed interest in name $j$.  We can thus
calculate the \emph{mean self-information} $I(\reco{k}{u})$ of the top
$k$ recommended names for every user $u$, yielding the
\emph{surprisal} of recommendation system \rec by averaging over all
users.
}

For assessing the prediction accuracy, we focus on the precision
scores. Please note that in this section, the impact of varying number
of known names on the prediction performance is not in the center of
interest and therefore precision scores are calculated for each user
on all but the known training items. Accordingly, the precision scores
decrease with increasing number of known names, as at the same time
the size of the corresponding test sets decreases.

\begin{figure}
  \centering
  \subfloat[][Personalization]{\label{fig:diversity:personalization-crop}
    \includegraphics[width=0.32\linewidth]{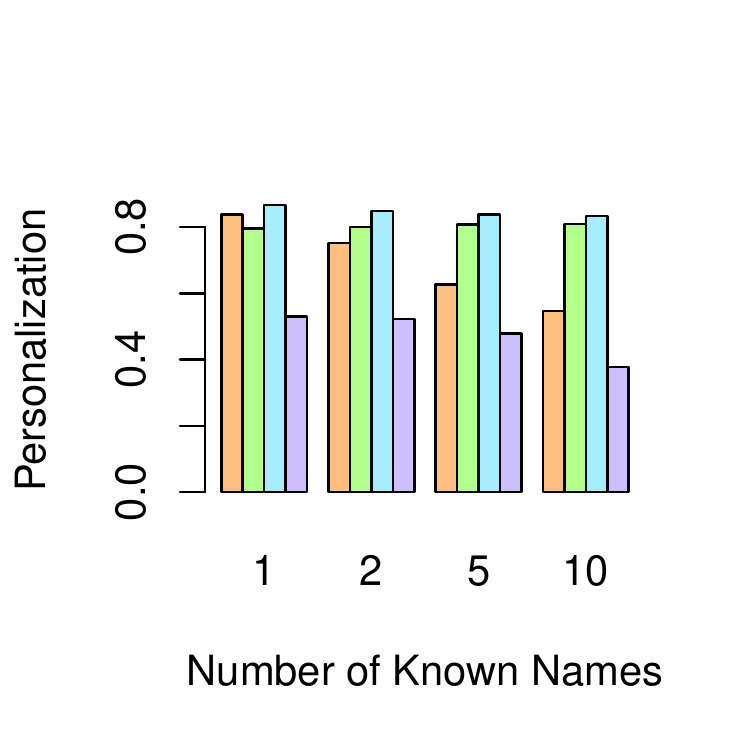}
  }
  \subfloat[][Precision]{\label{fig:diversity:precision-crop}
    \includegraphics[width=0.43\linewidth]{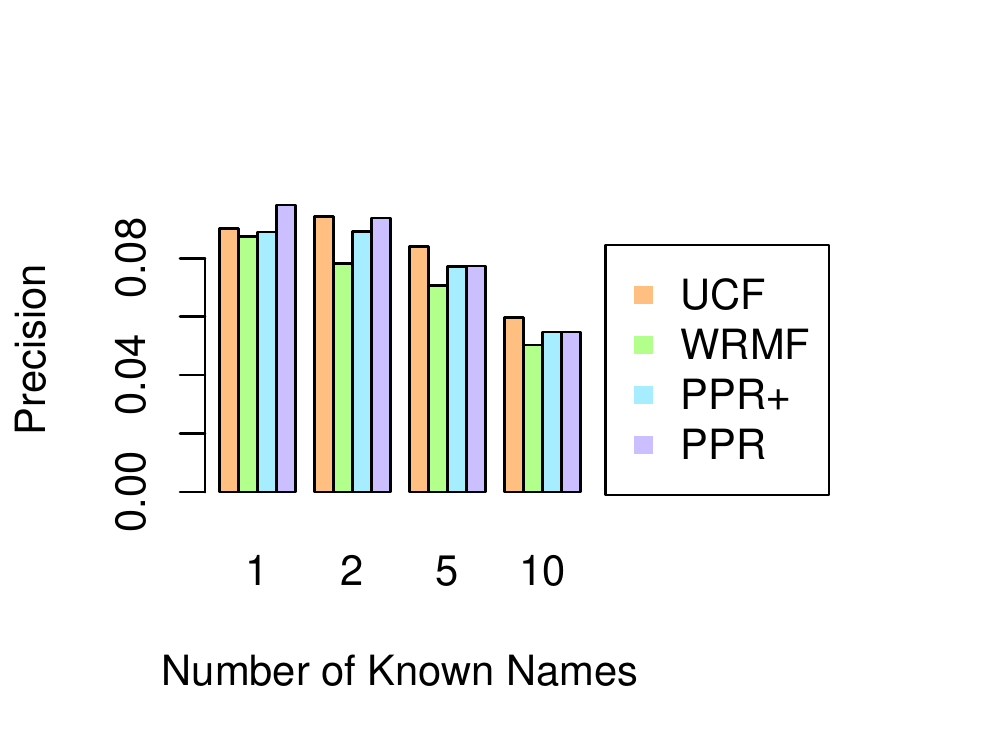}
  }
  \caption{Personalization as measure for assessing diversity and
    precision applied to the most promising recommendation systems
    from Sec.~\ref{sec:experiments:recommender}.}
  \label{fig:diversity:baseline}
\end{figure}

Fig.~\ref{fig:diversity:baseline} shows personalization and 
precision scores for the most promising recommendation systems from
Sec.~\ref{sec:recommender:experiments:results}, namely user based
collaborative filtering, weighted matrix factorization, \PPR and
\nameRank.
For user based collaborative filtering and \PPR, the personalization scores
indicate the lowest diversity of the recommendation results,
where the scores decrease with increasing number of known names. In
contrast, the results obtained from \WRMF and \nameRank are at a
respective constant level for the varying number of known names, at
which \nameRank achieves the highest scores. The low personalization
score for \PPR can be explained with the dominance of
global frequencies, \ie popular names. 

\makeatletter{}\subsection{Multi-Graph Aggregation for Increased Diversity}\enlargethispage*{2\baselineskip}
With respect to prediction accuracy and both considered measures for
diversity, our \nameRank approach already shows an superior trade-off
compared to the other considered recommendation systems. Nevertheless,
especially in the context of given names, diversity of recommendation
results is important, as future parents are using such a system in
order to find names which they like, but weren't aware of so far.

In the following we present and compare different approaches for
increasing the diversity of \nameRank. The evaluation of diversity in
an offline setting is very limited, as uncommon items (\ie names with
an high self-information score) are unlikely to be contained in an
user's test set. Also, diversity by itself can be maximized by
randomization, but at the same time, the results will show very low
accuracy with respect to the user's personal taste. 

We therefore compare the different diversification approaches by
considering the trade-off between prediction accuracy (in terms of
precision) and diversity (in terms of personalization). All proposed
diversification approaches encompass a diversification parameter
$\vec{\eta}$, which controls the impact of a second network of given
names, intended for introducing diversity to the training data.
Formally, we assume a set of graphs $G^{(1)}, \ldots, G^{(L)}$ with
$G^{(\ell)}\eqqcolon(V^{(\ell)}, E^{(\ell)})$ over a common vertex set
$V\supseteq V^{(\ell)}$ and edge weighting functions $w^{(1)}, \ldots,
w^{(L)}$ for $\ell=1,\ldots,L$.

\paragraph{Weighted Average}
As a baseline approach for combining ranking results obtained from
different networks, we consider the weighted average ranking \aveRank,
which is for given query items $\mathcal{I}\subseteq V$ defined by 
$
\aveRank(\mathcal{I})\coloneqq \frac{1}{L}\sum_{i=1}^L\eta_i\PPRa(\mathcal{I})
$
with $\eta_i\in[0,1]$ and $\sum_{i=1}^L\eta_i=1$.

We now look for ways of combining the graphs by themselves in order to
benefit from mutual reinforcement of important nodes during the
convergence process of the \nameRank algorithm. Various ways for
combining different graphs exist, of which we consider the following
two.

\begin{figure}
  \centering
  \subfloat[][Three separate graphs]{\label{fig:multigraph:separate}
    \includegraphics[width=0.32\linewidth]{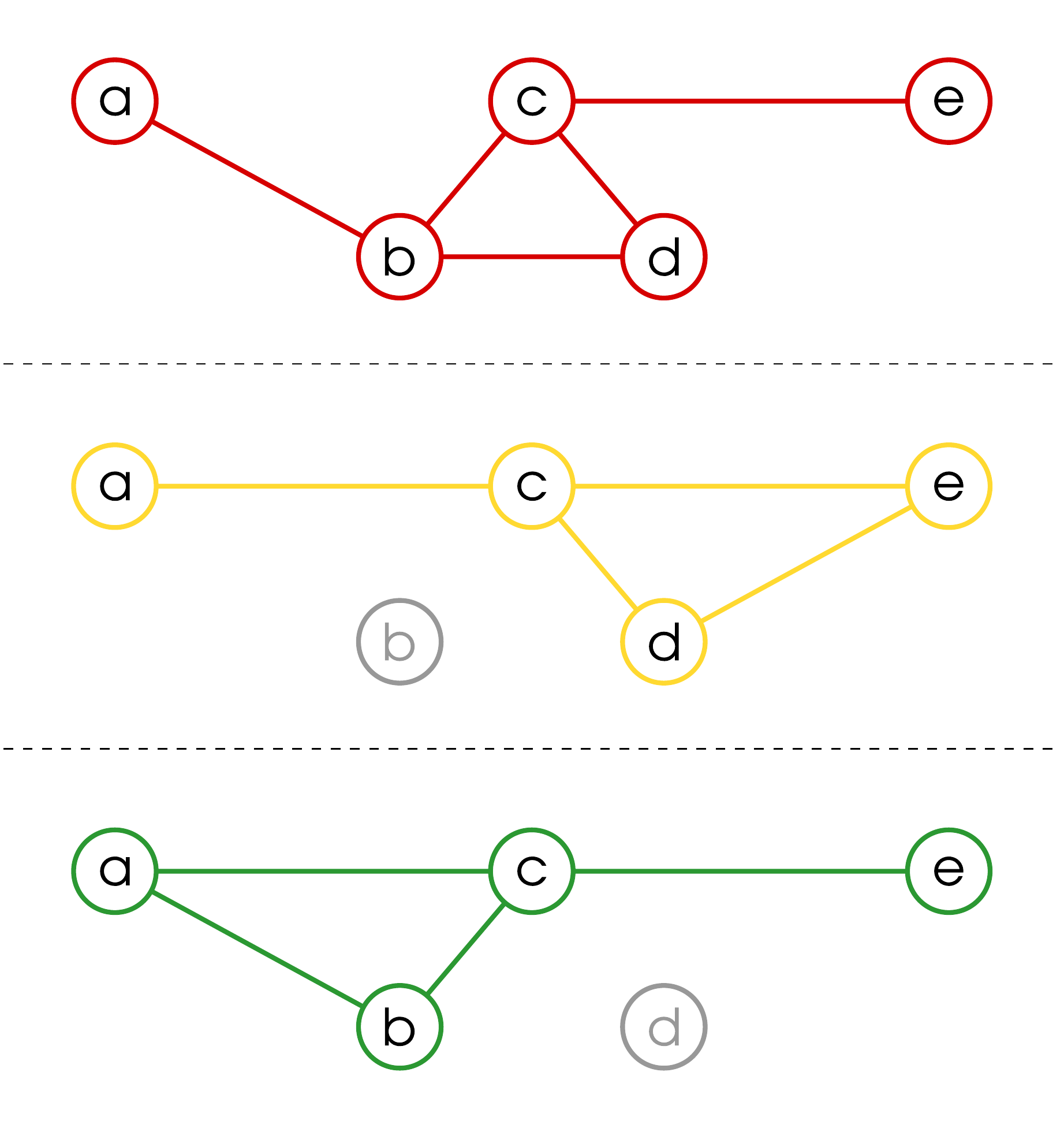}
  }
  \subfloat[][The merged multigraph]{\label{fig:multigraph:merged}
    \includegraphics[width=0.32\linewidth]{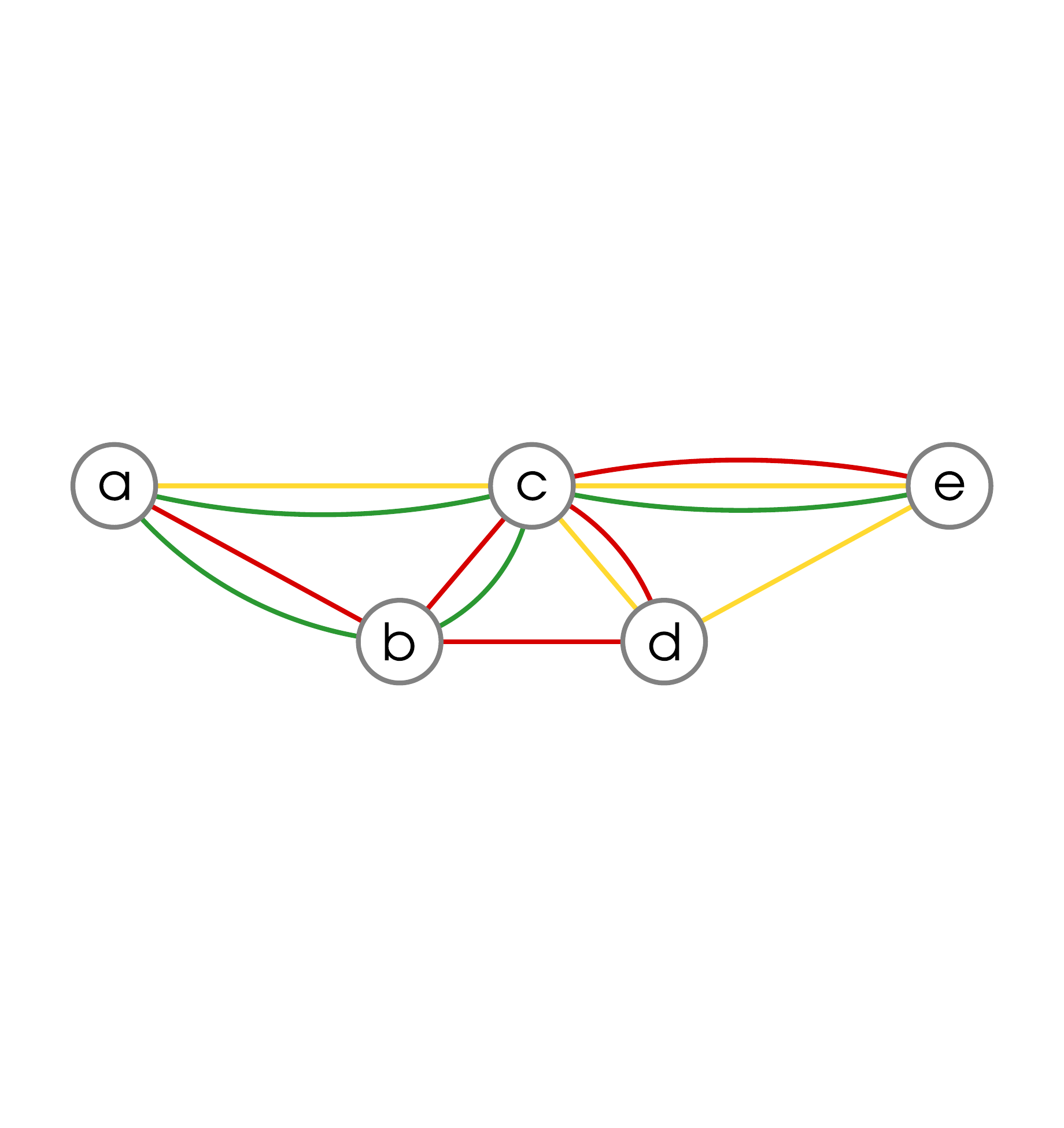}
  }
  \subfloat[][The combined graphs]{\label{fig:multigraph:combined}
    \includegraphics[width=0.32\linewidth]{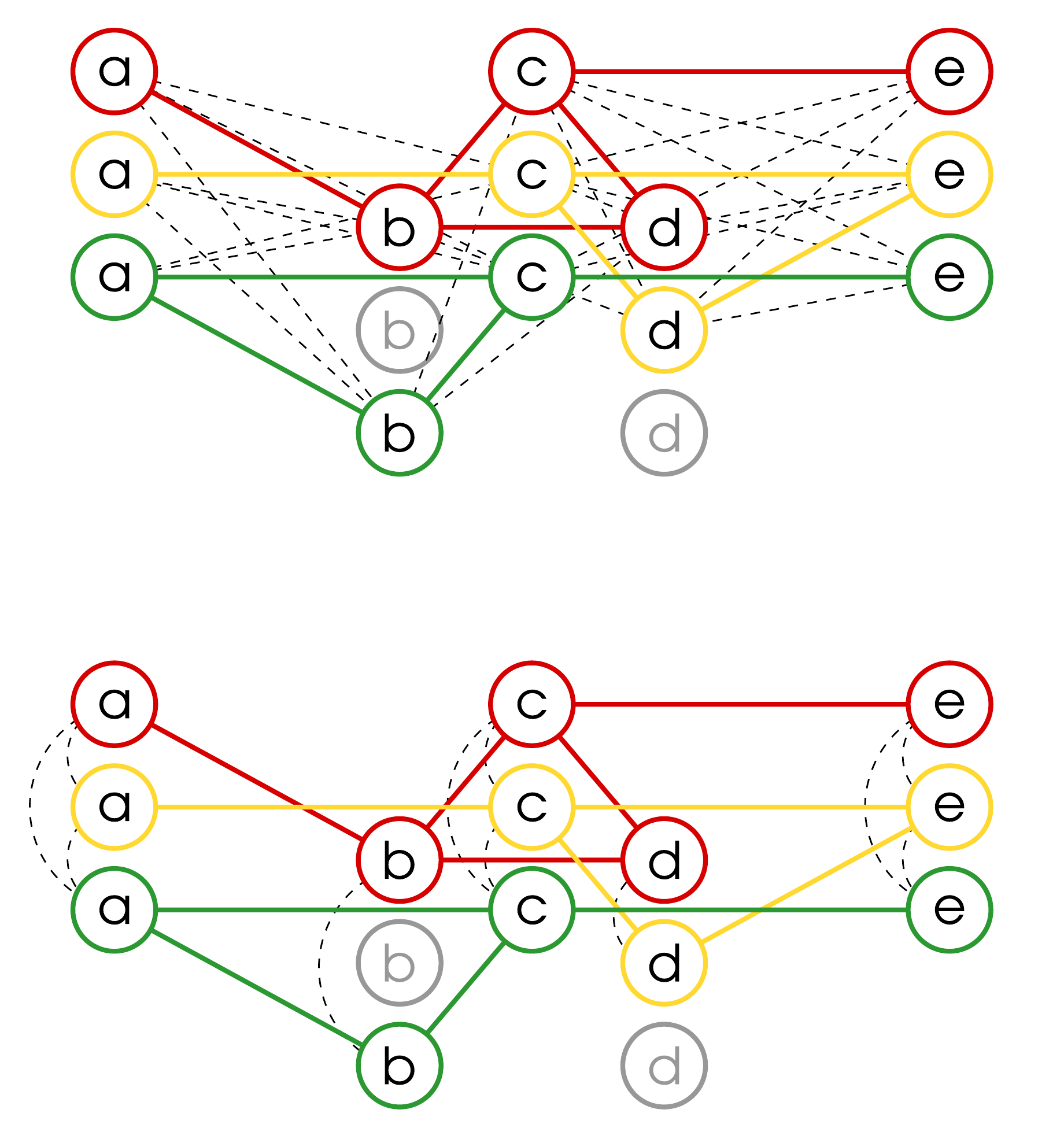}
  }
  \caption{Construction for the conditioned multigraph PageRank:
    Separate graphs (a), the merged multigraph (b) and the combined
    graphs, where nodes are duplicated according to the number of
    networks (c). The construction for the conditional multigraph
    \pagerank is shown on top and the for the parallel \pagerank on
    the bottom right.}
  \label{fig:multigraph:example}
\end{figure}

\paragraph{Conditional Multigraph PageRank}
The conditional multigraph PageRank is based on the idea of a
multigraph built from $G^{(1)},\ldots,G^{(L)}$ in which edges are
additionally labeled corresponding the graph they originated from.
Accordingly, multiple edges among vertex pairs may exist, as depicted
in Fig.~\ref{fig:multigraph:merged}. Applying the intuition of the
random surfer model~\todo{REFERENZ}, the basic idea of the conditional
multigraph PageRank is, that a random surfer, reaching node $u$ via a
certain edge type, will more likely leave $u$ using an edge of the
same type, but may nevertheless leave $u$ using any other incident
edge. Thus, the importance of each graph's neighborhood is accented,
but nevertheless, interrelations with other nodes as induced by other
graphs are also considered. Furthermore, the transition probability of
an edge $(u,v)$ scales with the number of graphs containing it. By
introducing according stochastic\footnote{\ie rows and columns sum up
  to one} damping factors
$(\boldsymbol{\eta})_{\ell_1,\ell_2}\in[0,1]^{L\times L}$, the overall
inter-graph transition probabilities can be controlled.

In  order  to  apply  standard  \pagerank  calculations,  we  build  a
\emph{combined}   (non-multi)   graph   $\widetilde{G}=(\widetilde{V},
\widetilde{E})$,   which  incorporates  the   conditional  multi-graph
approach  by construction. For  this, we  firstly duplicate  each node
$v\in    V$    according    to    the   number    of    graphs,    \ie
$\widetilde{V}\coloneqq{v^{1}_1,\ldots,                 v^{1}_n,\ldots,
  v^{L}_1,\ldots,   v^{L}_n}$.  Secondly,   we   include  each   graph
$G_\ell$'s connectivity  within $\widetilde{E}_{\ell,\ell}$, resulting
in           $\widetilde{E}_{\ell,\ell}\coloneqq\set{(u^\ell,v^\ell)\in
  \widetilde{V}\times\widetilde{V}\mid (u,v)\in E_\ell}$. Finally, for
$\ell_1\ne\ell_2$,                        we                       set
$\widetilde{E}_{\ell_1,\ell_2}\coloneqq\set{(u^{\ell_1},
  v^{\ell_2})\mid   (u^{\ell_2},   v^{\ell_2})\in   E_{\ell_2}}$   and
$\widetilde{E}\coloneqq\bigcup\limits_{\ell_1,\ell_2}\widetilde{E}_{\ell_1,\ell_2}$.
Edge   weights  are   given   by  the   combined  weighting   function
$\widetilde{w}(u^{\ell_1},v^{\ell_2})\coloneqq
w_{\ell_2}(u^{\ell_1},v^{\ell_2})$.   Accordingly,   $\widetilde{G}$'s
adjacency matrix $A$ is block-wise structured
\[ 
A= 
\left(
\begin{BMAT}[8pt]{c:c:c}{c:c:c}
  A^{1,1} & \ldots & A^{1,L} \\
  \vdots & \ddots & \vdots \\
  A^{L,1} & \ldots & A^{L,L}
\end{BMAT} 
\right)
\]
where each $n\times n$ sub-matrix $A^{\ell,\ell}$ corresponds to graph
$\ell$'s adjacency matrix and $A^{\ell_1,\ell_2}$ describes
transitions from graph $\ell_1$ to graph $\ell_2$. For calculating
\pagerank scores on $\widetilde{G}$, its adjacency matrix $A$ has to
be normalized in order to be column-stochastic, \ie all columns sum up
to one. The accentuation of inner-graph connectivity is realized by
introducing damping factors $\eta_{\ell_1,\ell_2}$ on the graph
normalization, such that each of $A^{\ell_1,\ell_2}$'s columns sums up
to $\eta_{\ell_1,\ell_2}$. For obtaining the conditional multigraph
\pagerank score for a given $\mathcal{I}$, we calculate
$\multiRank(\mathcal{I})\coloneqq\PPRa(\mathcal{I})$ on $\widetilde{G}$.

\paragraph{Parallel \pagerank}
A simpler combination of $G^{(1)},\ldots,G^{(L)}$ can be achieved, by
interconnecting each node $v\in V$ across all graphs. Fixing the
notation introduced for the conditional multigraph \pagerank, we build
an according graph $\widetilde{G}=(\widetilde{V}, \widetilde{E})$ by
altering only the inter-graph connectivity to
$\widetilde{E}_{\ell_1,\ell_2}\coloneqq\set{(u^{\ell_1},u^{\ell_2})\mid
  u\in V}$. For obtaining the parallel multigraph \pagerank score for
a given $\mathcal{I}$, we calculate
$\pRank(\mathcal{I})\coloneqq\PPRa(\mathcal{I})$ on $\widetilde{G}$.

\paragraph{Weighting}
Please note that the prediction performance of \PPRa depends on the
edge weighting of the input graph. Accordingly, the weights of
inter-network edges may be adopted in order to improve one of the
evaluation targets (\eg depending of the incident node's
connectivity).  For our evaluation, we set the weight of every such edge
$(u^{\ell_1},v^{\ell_2})$ with $\ell_1\ne\ell_2$ to one in order to
consider the most general case.

\makeatletter{}\subsection{Evaluation}
For evaluating the different result diversification approaches for
\PPRa, we consider the network of given names obtained from
\nameling's usage data and combine it pairwise with the co-occurrence
graph obtained from \wikipedia and respectively \twitter, as well as
the shared category graph obtained from \wiktionary and the shared
favorite graph from \nameling. We thereby apply the evaluation
protocol from Sec.~\ref{sec:recommender:experiments:results}, except
that all but the training names are used for evaluation. For
considering the trade-off between \emph{prediction accuracy} and
\emph{diversity}, we show the obtained results in a
precision/diversity scatter plot. For reference, we also calculated
the diversity of the considered user profiles by themselves ($h=0.99$,
$I=9.62$, calculated on the first ten test items for $k=1,5$).

\begin{figure}
  \centering
  \includegraphics[scale=0.4]{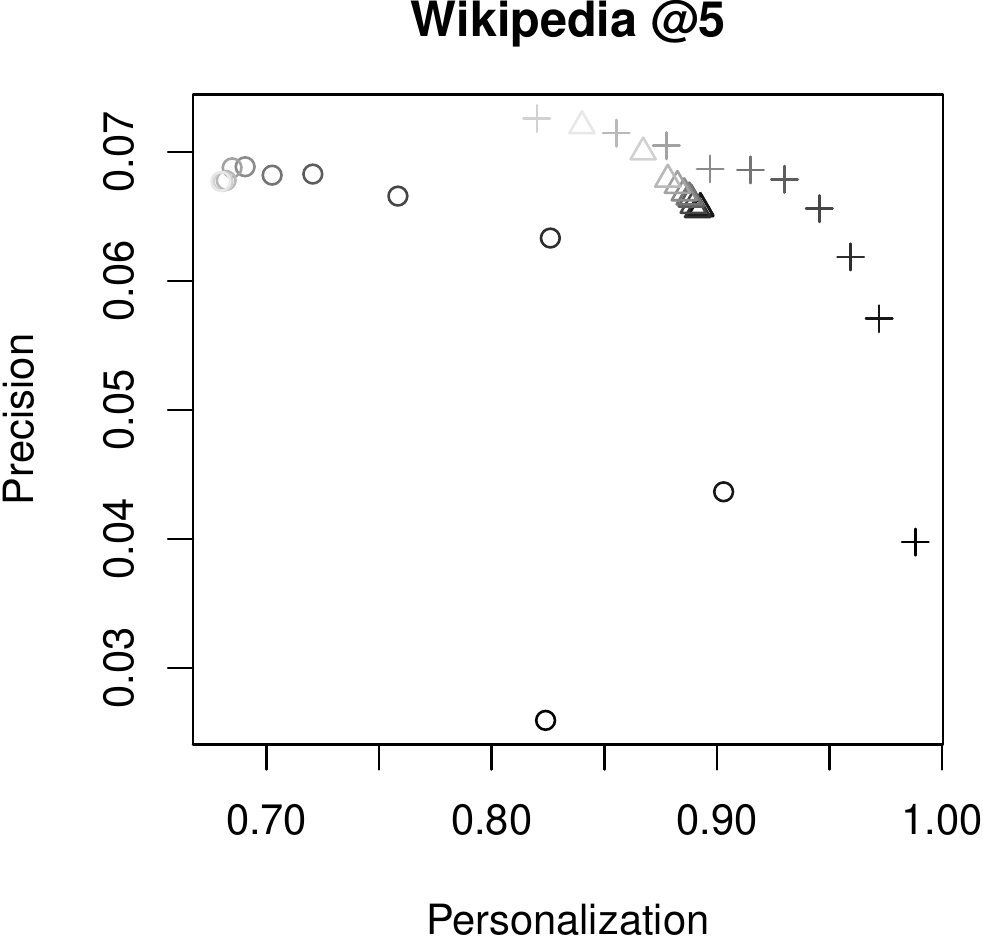}\quad\quad\quad
  \includegraphics[scale=0.4]{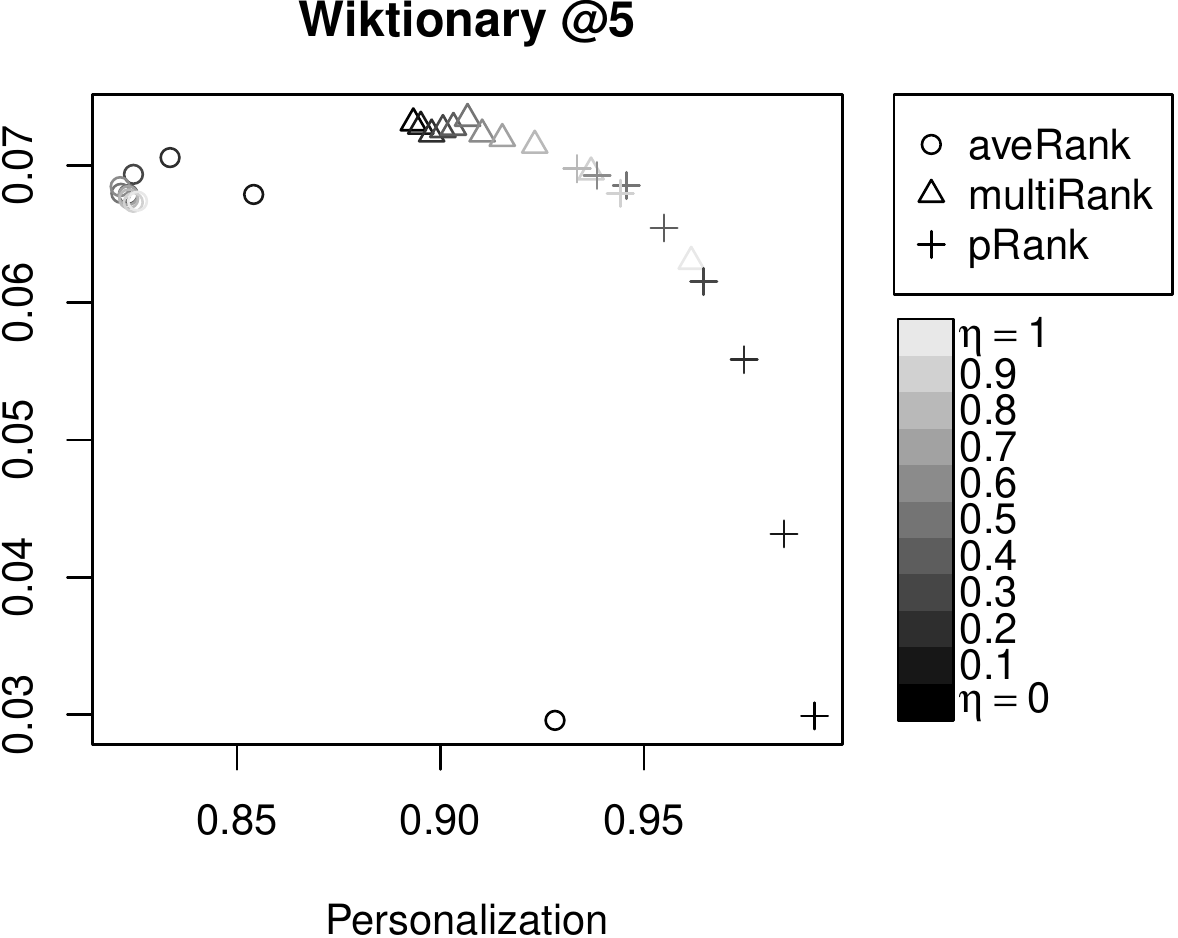}\\
  \includegraphics[scale=0.4]{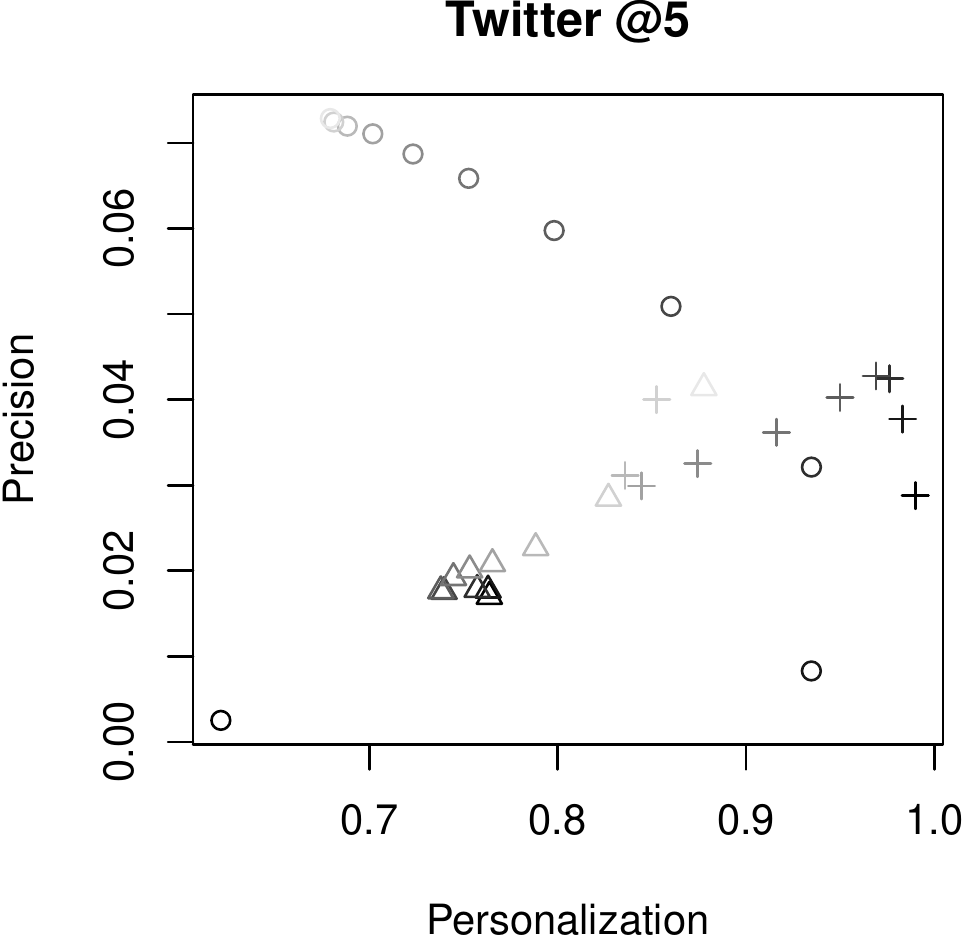}\quad\quad\quad
  \includegraphics[scale=0.4]{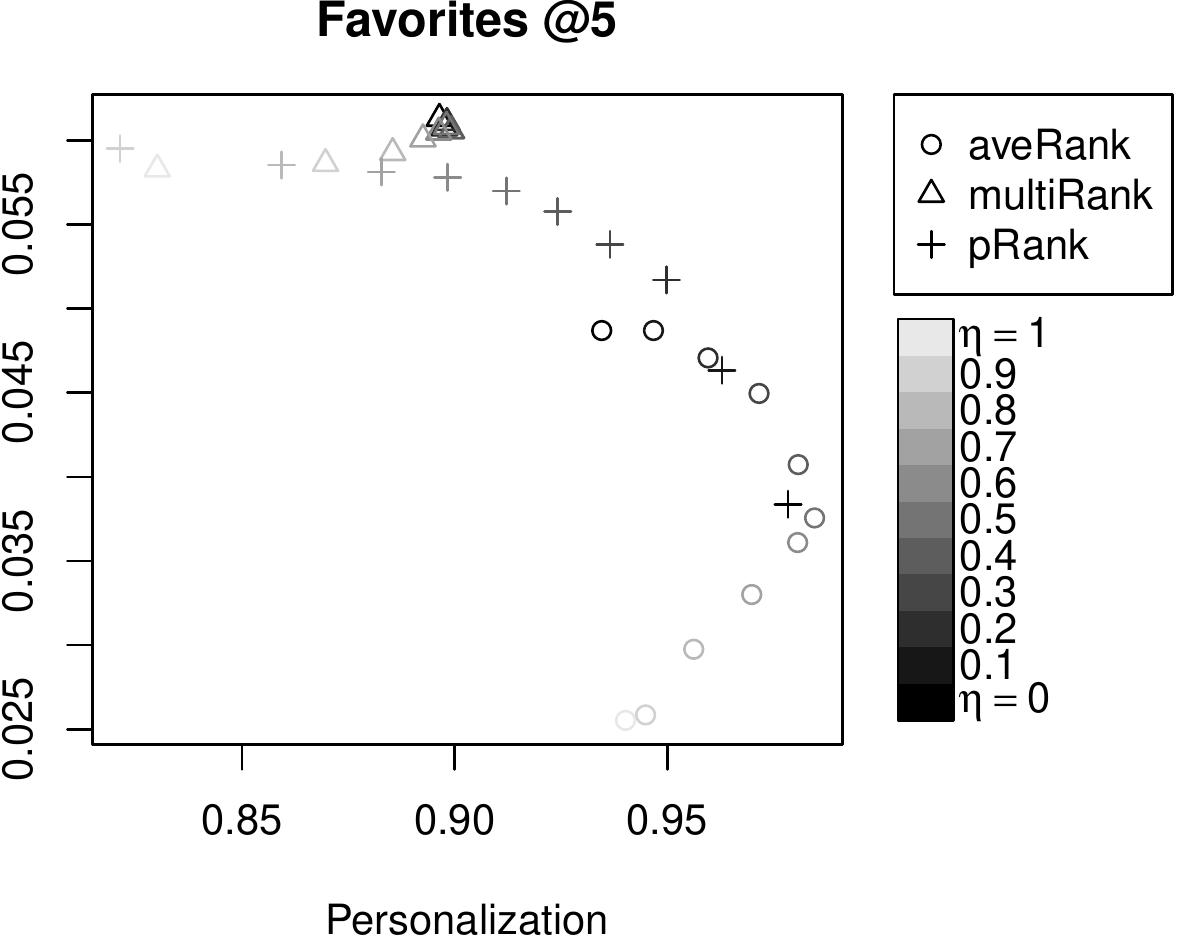}\\
\comments{
  \includegraphics[scale=0.38]{personalization-tweets_jure-1-crop}
  \includegraphics[scale=0.38]{personalization-tweets_jure-5-crop}
  \includegraphics[scale=0.38]{personalization-tweets_jure-10-crop}\\
  \includegraphics[scale=0.38]{personalization-vornamen_lc-1-crop}
  \includegraphics[scale=0.38]{personalization-vornamen_lc-5-crop}
  \includegraphics[scale=0.38]{personalization-vornamen_lc-10-crop}\\
  \includegraphics[scale=0.38]{personalization-shared_name-categories-1-crop}
  \includegraphics[scale=0.38]{personalization-shared_name-categories-5-crop}
  \includegraphics[scale=0.38]{personalization-shared_name-categories-10-crop}\\
  \includegraphics[scale=0.38]{personalization-shared_name-favorite-1-crop}
  \includegraphics[scale=0.38]{personalization-shared_name-favorite-5-crop}
  \includegraphics[scale=0.38]{personalization-shared_name-favorite-10-crop}\\
}
  \caption{Diversity vs. accuracy scatter plots for all considered
    multigraph approaches in the co-occurrence network obtained from
    \twitter (top row), the co-occurrence network obtained from
    \wikipedia (2nd row), the shared categories network (3rd row) and
    shared favorite names network (bottom row) for recommendations
    based on one, five or ten known names in the left, middle and
    right column respectively.}
  \label{fig:diversity:evaluation}
\end{figure}

Fig.~\ref{fig:diversity:evaluation} shows the accuracy/diversity
scatter plots with respect to personalization on all considered pairs
of networks. We only show results obtained for the TakeFirstIn
protocol with $k=5$. The results only differ in magnitude for varying
$k$ but the overall characteristics are unchanged.  We note that all
considered multigraph approaches introduce varying amounts of
diversity and an accordingly differing level of accuracy.\enlargethispage*{2\baselineskip}
We firstly consider the personalization/precision trade-off. The
weighted average approach shows similar characteristics for all
considered networks. Except for the combination with the \twitter
based network, both combined multigraph approaches outperform the
considered baseline, where \pRank shows overall smother transitions
and covers broader ranges of the personalization metric. Nevertheless,
\multiRank is the only approach which shows increasing tendencies for
the precision scores with increasing personalization score. In
combination with the \twitter based network, both combined multigraph
approaches show inferior performance than the baseline model.

Summing up, we observe differing characteristics for the
accuracy/diversity lemma, hinting at a more pronounced dependence of
\multiRank on the inter-network correlations. Nevertheless, the
results obtained from the different approaches are complementary and
ultimately only a live evaluation of the different graph combinations will
reveal, which result diversification approach corresponds best to the
user's expectations.

\comments{
\begin{figure}
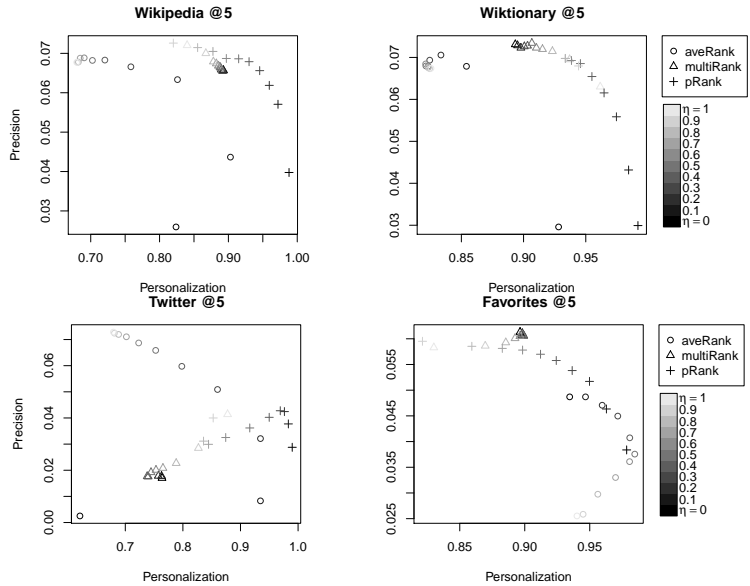

  \centering
  \includegraphics[scale=0.44]{surprisal-tweets_jure-1-crop}
  \includegraphics[scale=0.44]{surprisal-tweets_jure-5-crop}
  \includegraphics[scale=0.44]{surprisal-tweets_jure-10-crop}\\
  \includegraphics[scale=0.44]{surprisal-vornamen_lc-1-crop}
  \includegraphics[scale=0.44]{surprisal-vornamen_lc-5-crop}
  \includegraphics[scale=0.44]{surprisal-vornamen_lc-10-crop}\\
  \includegraphics[scale=0.44]{surprisal-shared_name-categories-1-crop}
  \includegraphics[scale=0.44]{surprisal-shared_name-categories-5-crop}
  \includegraphics[scale=0.44]{surprisal-shared_name-categories-10-crop}\\
  \includegraphics[scale=0.44]{surprisal-shared_name-favorite-1-crop}
  \includegraphics[scale=0.44]{surprisal-shared_name-favorite-5-crop}
  \includegraphics[scale=0.44]{surprisal-shared_name-favorite-10-crop}\\
  \caption{Diversity vs. accuracy scatter plots for all considered
    multigraph approaches in the co-occurrence network obtained from
    \twitter (top row), the co-occurrence network obtained from
    \wikipedia (2nd row), the shared categories network (3rd row) and
    shared favorite names network (bottom row) for recommendations
    based on one, five or ten known names in the left, middle and
    right column respectively.}
  \label{fig:diversity:evaluation:surprisal}
\end{figure}
}

            \makeatletter{}\section{Related Work}\label{sec:related}
\label{names:related}
\comments{
  * KEYWORDS: distributional semantics, corpora analysis, semantic relatedness

  * named entity disambiguation
  * person name vs. given name
  * 
  * ESA, WikiRelate

  * geographic co-occurrences

  CONTRIBUTION
  -> focus on evaluation of performance of building blocks
  ->

  ``Using Encyclopedic Knowledge for Named Entity Disambiguation''
  -> detect entities -> disambiguation using
     word contexts -> similarities

  ``WikiRelate''

  ``ESA''

  ``Large-Scale Named Entity Disambiguation Based on Wikipedia Data''

==== Statistical semantics
 - Second Order Co-occurrence PMI for Determining the Semantic
 Similarity of Words

 - 

==== Semantic similarity or semantic relatedness

==== Link Prediction / Vertex Similarity

}The present work introduces a new field of application for analyzing
relations among named entities and recommendation systems. It is
motivated by the work on the search engine ``\nameling'', which is
presented in~\cite{mitzlaff2012namelings}. The ranking performance of
the structural similarity metrics based on the \wikipedia corpus
relative to the actual usage data which accrued in the running system
is evaluated in~\cite{mitzlaff2012ranking,mitzlaff2012relatedness}. The
considered approaches are based on work in the field of
\emph{distributional semantics}, \emph{link prediction} and (more generally) \emph{vertex
  similarity} in graphs as well as \emph{recommendation systems}.

\paragraph{Distributional Semantics}
The field of distributional semantics relatedness has attracted a lot
of attention in literature during the past decades (see
\cite{cohen2009empirical} for a review). Several statistical measures
for assessing the similarity of words are proposed, as for example
in~\cite{lesk1969word,grefenstette1992finding,islam2006second,landauer1997solution,turney2001mining}. Notably,
first approaches for using \wikipedia as a source for discovering
relatedness of concepts can be found
in~\cite{bunescu2006using,strube2006wikirelate,gabrilovich2007computing}.

\paragraph{Vertex Similarity \& Link Prediction}
In the context of social networks, the task of predicting (future)
links is especially relevant for online social networks, where social
interaction is significantly stimulated by suggesting people as
contacts which the user might know. From a methodological point of
view, most approaches build on different similarity metrics on pairs
of nodes within weighted or unweighted
graphs~\cite{jeh2002simrank,leicht2005vertex,lu2009similarity,lu2010prediction}. A
good comparative evaluation of different similarity metrics is
presented in \cite{libennowell2007linkprediction}. The construction of
multigraph extensions to the NameRank algorithm for controlled result
diversification are based on~\cite{scholz2012insights}, where random
walks are used to combine the information of different networks.

\paragraph{Recommender Systems}
Recommending given names is just a special case for the
\emph{item recommendation} task which is extensively discussed for
various fields of application, such as
movies~\cite{golbeck2006filmtrust}, tags~\cite{jaschke2008tag} and
products~\cite{linden2003amazon}. Notably the characteristics of
deriving recommendations from activity logs (rather than explicit user
feedback) are discussed in~\cite{hu2008collaborative} where the
uncertainty of implicit feedback taken into account, by attaching an
adequate level of confidence to an implicitly expressed interest in an
item (\eg by clicking on the description of a given name). A very good
introduction and summary of the evaluation of recommendation systems
can be found in~\cite{shani2011evaluating}.

            \bibliographystyle{abbrv}
   \bibliography{bibliography}

\end{document}